\documentclass{emulateapj}
\pdfoutput=1
\usepackage{graphicx}
\usepackage[space]{grffile}
\usepackage{latexsym}
\usepackage{textcomp}
\usepackage{longtable}
\usepackage{multirow,booktabs}
\usepackage{amsfonts,amsmath,amssymb}
\usepackage{natbib}
\usepackage{url}
\usepackage{CJKutf8}
\usepackage{hyperref}
\hypersetup{colorlinks=false,pdfborder={0 0 0}}
\newif\iflatexml\latexmlfalse
\usepackage[utf8]{inputenc}
\usepackage[english]{babel}

\newcommand\T{\rule{0pt}{2.6ex}}       
\newcommand{\vsini}{$v$~sin~$i$}

\shorttitle{Data Release 13}
\shortauthors{SDSS Collaboration}

\begin{document}


\title{The Thirteenth Data Release of the Sloan Digital Sky Survey: First Spectroscopic Data from the SDSS-IV Survey MApping Nearby Galaxies at Apache Point Observatory}


\author{Franco~D.~Albareti\altaffilmark{1,2,3},
Carlos~Allende~Prieto\altaffilmark{4,5},
Andres~Almeida\altaffilmark{6},
Friedrich~Anders\altaffilmark{7},
Scott~Anderson\altaffilmark{8},
Brett~H.~Andrews\altaffilmark{9},
Alfonso~Arag{\'o}n-Salamanca\altaffilmark{10},
Maria~Argudo-Fern{\'a}ndez\altaffilmark{11,12},
Eric~Armengaud\altaffilmark{13},
Eric~Aubourg\altaffilmark{14}
% ...
}
\altaffiltext{1}{Instituto de F\'{\i}sica Te\'orica, (UAM/CSIC), Universidad Aut\'onoma de Madrid, Cantoblanco, E-28049 Madrid, Spain}
\altaffiltext{2}{Campus of International Excellence UAM+CSIC, Cantoblanco, E-28049 Madrid, Spain}
\altaffiltext{3}{la Caixa-Severo Ochoa Scholar}
\altaffiltext{4}{Instituto de Astrof\'{\i}sica de Canarias, E-38205 La Laguna, Tenerife, Spain}
\altaffiltext{5}{Departamento de Astrof\'{\i}sica, Universidad de La Laguna (ULL), E-38206 La Laguna, Tenerife, Spain}
\altaffiltext{6}{Instituto de Investigaci\'on Multidisciplinario en Ciencia y Tecnolog\'{\i}a, Universidad de La Serena, Benavente 980, La Serena, Chile}
\altaffiltext{7}{Leibniz-Institut fuer Astrophysik Potsdam (AIP), An der Sternwarte 16, D-14482 Potsdam, Germany}
\altaffiltext{8}{Department of Astronomy, University of Washington, Box 351580, Seattle, WA 98195, USA}
\altaffiltext{9}{PITT PACC, Department of Physics and Astronomy, University of Pittsburgh, Pittsburgh, PA 15260, USA}
\altaffiltext{10}{School of Physics and Astronomy, University of Nottingham, University Park, Nottingham, NG7 2RD, UK}
\altaffiltext{11}{Unidad de Astronom\'{\i}a, Universidad de Antofagasta, Avenida Angamos 601, Antofagasta 1270300, Chile}
\altaffiltext{12}{Shanghai Astronomical Observatory, Chinese Academy of Science, 80 Nandan Road, Shanghai 200030, P. R. China}
\altaffiltext{13}{IRFU, CEA, Centre d'Etudes Saclay, 91191 Gif-Sur-Yvette Cedex, France}
\altaffiltext{14}{APC, University of Paris Diderot, CNRS/IN2P3, CEA/IRFU, Observatoire de Paris, Sorbonne Paris Cite, France}
% ...

\begin{abstract}
The fourth generation of the Sloan Digital Sky Survey (SDSS-IV) began observations in July 2014. It pursues three core programs: the Apache Point Observatory Galactic Evolution Experiment 2 (APOGEE-2), Mapping Nearby Galaxies at APO (MaNGA), and the Extended Baryon Oscillation Spectroscopic Survey (eBOSS). As well as its core program, eBOSS contains two major  subprograms:  the Time Domain Spectroscopic Survey (TDSS) and the SPectroscopic IDentification of ERosita Sources (SPIDERS). This paper describes the first data release from SDSS-IV, Data Release 13 (DR13). DR13 makes publicly available the first 1390 spatially resolved integral field unit observations of nearby galaxies from MaNGA. It includes new observations from eBOSS, completing the Sloan Extended QUasar, Emission-line galaxy, Luminous red galaxy Survey (SEQUELS), which also targeted variability-selected objects and X-ray selected objects. DR13 includes new reductions of the SDSS-III BOSS data, improving the spectrophotometric calibration and redshift classification, and new reductions of the SDSS-III APOGEE-1 data, improving stellar parameters for dwarf stars and cooler stars. DR13 provides more robust and precise photometric calibrations. Value-added target catalogs relevant for eBOSS, TDSS, and SPIDERS and an updated red-clump catalog for APOGEE are also available. This paper describes the location and format of the data and provides references to important technical papers. The SDSS website, www.sdss.org, provides links to the data, tutorials, examples of data access, and extensive documentation of the reduction and analysis procedures. DR13 is the first of a scheduled set that will contain new data and analyses from the planned $\sim$ 6-year operations of SDSS-IV.
\end{abstract}


\section{Introduction}
\label{sec:intro}
The {\bf S}loan {\bf D}igital {\bf S}ky {\bf S}urvey (SDSS) has been observing the Universe using the 2.5-meter Sloan Foundation Telescope \citep{2006AJ....131.2332G} at {\bf A}pache {\bf P}oint {\bf O}bservatory (APO) for over 15 years. The goal of the original survey \citep[2000--2005;][]{york2000} was to map large-scale structure with five-band imaging over $\sim \pi$ steradians of the sky and spectra of $\sim 10^6$ galaxies and $\sim$10$^5$ quasars. This program was accomplished using a drift-scan camera \citep{1998AJ....116.3040G} and two fiber-fed optical R$\sim$1800 spectrographs \citep{2013AJ....146...32S}, each with 320 fibers. 

The imaging and spectroscopy goals were not entirely fulfilled in the initial five-year period, and thus SDSS-I was followed by SDSS-II \citep[2005--2008;][]{2009ApJS..182..543A}. Its first goal was to complete the planned initial large scale structure redshift survey as the Legacy program. It added SEGUE \citep[{\bf S}loan {\bf E}xtension for {\bf G}alactic {\bf U}nderstanding and {\bf E}xploration;][]{2009AJ....137.4377Y}, a spectroscopic survey focused on stars, and imaged an average of once every five days a $\sim$ 200 sq. deg area along the celestial equator with repeated scans in SDSS-I (``Stripe 82''), to search for Type Ia supernovae and other transients \citep{2008AJ....135..338F}. 

The success of SDSS as a cosmological probe, particularly the detection of the clustering of luminous red galaxies (LRG) on the 100 $h^{-1}$ Mpc scale expected from baryon acoustic oscillations \citep[BAO;][]{2005ApJ...633..560E}, led to the conception and implementation of BOSS \citep[{\bf B}aryon {\bf O}scillation {\bf S}pectroscopic {\bf S}urvey;][]{2013AJ....145...10D} as the flagship program in the third version of the survey, SDSS-III \citep[2008--2014;][]{2011AJ....142...72E}. As part of BOSS, SDSS-III imaged additional areas in the part of the south Galactic cap visible from the Northern hemisphere. At the conclusion of these observations, the SDSS imaging camera was retired and is now part of the permanent collection of the Smithsonian National Air and Space Museum\footnotemark[1]\footnotetext[1]{\url{https://airandspace.si.edu/collection-objects/camera-imaging-digital-sloan-digital-sky-survey-ccd-array}}. During the summer shutdown in 2009, the original SDSS spectrographs were replaced by new, more efficient, spectrographs to be used by BOSS. The BOSS spectrographs featured expanded wavelength coverage (3560\AA$<\lambda<$10400 \AA), new CCD detectors with improved read noise, smaller pixels (15$\mu$m), and improved quantum efficiency, and VPH gratings instead of the original replicated surface relief gratings \citep{2013AJ....146...32S}. The two spectrographs were now fed by 500 fibers each so that the desired number of redshifts could be reached in the planned survey lifetime. 

During the first year of SDSS-III (2008--2009), the SEGUE-2 survey (Rockosi et al., in preparation) used the original SDSS spectrographs to observe additional Milky Way halo fields to target distant halo samplers and trace substructure. In SDSS-III all bright time could be used for scientific observations with the arrival of two new instruments. MARVELS \citep[{\bf M}ulti-object {\bf A}PO {\bf R}adial {\bf V}elocity {\bf E}xoplanet {\bf L}arge-area {\bf S}urvey;][] {2015AJ....149..186P, 2016PASP..128d5003T} used a novel multiplexing interferometer to observe 60 stars simultaneously to search for radial velocity variations caused by hot Jupiters and close brown dwarf companions. APOGEE \citep[{\bf A}pache {\bf P}oint {\bf O}bservatory {\bf G}alactic {\bf E}volution \textit{E}xperiment;][]{Majewski_2017} used a 300-fiber, R$\sim$22,000 H-band spectrograph (\citealt{Majewski_2017}; Wilson et al., in preparation) to measure stellar parameters, chemical abundances, and radial velocities, mainly for red giants \citep{zasowski2013}.  

Since routine operations started in 2000, there have been thirteen public data releases. All data releases are cumulative, re-releasing the best reduction of all previously taken data. The most recent of these was Data Release 12 \citep{2015ApJS..219...12A}, which contained all of the SDSS-III data, as well as the re-reduced data from SDSS-I and SDSS-II. SDSS-I to SDSS-III imaged 14,555 degree$^2$ in the five filters \citep{1996AJ....111.1748F,2010AJ....139.1628D}. Most of the sky was surveyed once or twice, but regions in Stripe 82 were observed between 70 and 90 times. By the time of their retirement, the SDSS spectrographs had obtained R$\sim$1800 optical spectra for 860,836 galaxies, 116,003 quasars, and 521,990 stars. With the BOSS spectrographs, the survey has added data with similar resolution for 1,372,737 galaxies, 294,512 quasars, and 247,216 stars.  APOGEE has contributed high-resolution IR spectra of 156,593 stars. MARVELS had observed 3233 stars with at least 16 epochs of radial velocity measurements.  

The success of the previous Sloan Digital Sky Surveys and the continuing importance of the wide-field, multiplexing capability of the Sloan Foundation Telescope motivated the organization of the fourth phase of the survey, SDSS-IV \citep{overview}. SDSS-IV extends SDSS observations to many fibers covering the spatial extent of nearby galaxies, to new redshift regimes, and to the parts of the Milky Way and dwarf galaxies that are only visible from the Southern Hemisphere. The MaNGA ({\bf Ma}pping {\bf N}earby {\bf G}alaxies at {\bf A}PO) survey studies galaxy formation and evolution across a wide range of masses and morphological types by observing a substantial portion of the optical spatial extent of $\sim$10$^4$ galaxies \citep{Bundy_2015}. It accomplishes this goal by employing 17 bundles ranging in size between 19 and 127 fibers to cover targets selected from an extended version of the NASA-Sloan Atlas\footnotemark[2]\footnotetext[2]{\url{http://www.nsatlas.org}} and 12 bundles of 7 fibers for calibration stars. These integral field units (IFUs) feed the BOSS spectrographs, providing information on the properties of gas and stars in galaxies out to 1.5--2.5 effective radii ($R_e$). 

Another survey, eBOSS \citep[{\bf e}xtended {\bf B}aryon {\bf O}scillation {\bf S}pectroscopic {\bf S}urvey][]{2016AJ....151...44D}, shares the dark time equally with MaNGA. eBOSS will measure with percent-level precision the distance-redshift relation with BAO in the clustering of matter over the relatively unconstrained redshift range $0.6 < z < 2.2$. This redshift range probes the Universe during its transition from matter-dominated to dark-energy-dominated. Multiple measurements of the angular diameter distance ($d_A(z)$) and Hubble parameter ($H(z)$) from BAO over the redshifts covered by eBOSS are therefore crucial for understanding the nature of dark energy. eBOSS will use spectroscopic redshifts from more than 400,000 LRGs and nearly 200,000 Emission-Line Galaxies (ELGs) to extend the final BOSS galaxy clustering measurements \citep{alam2016} by providing two new BAO distance measurements over  the redshift interval $0.6< z <1.1$  Roughly 500,000 spectroscopically-confirmed quasars will be used as tracers of the underlying matter density field at $0.9 < z < 2.2$, providing the first measurements of BAO in this redshift interval. Finally, the Lyman-$\alpha$ forest imprinted on approximately 120,000 new quasar spectra will give eBOSS an improved BAO measurement over that achieved by BOSS \citep{delubac2015,bautista2017}. The three new tracers will provide BAO distance measurements with a precision of 1\% at $z=0.7$ (LRG), 2\% at $z=0.85$ (ELG), and 2\% at $z=1.5$ (quasar) while the enhanced Lyman-$\alpha$ forest sample will improve BOSS constraints by a factor of 1.4. Furthermore, the clustering from eBOSS tracers will allow new measurements of redshift-space distortions (RSD), non-Gaussianity in the primordial density field, and the summed mass of neutrino species.  Extensively observing these redshift ranges for the first time in SDSS required re-evaluation of targeting strategies. Preliminary targeting schemes for many of these classes of objects were tested as part of SEQUELS ({\it S}loan {\it E}xtended {\it QU}asar, {\it E}mission-line galaxy, {\it L}uminous red galaxy {\it S}urvey), which used 126 plates observed across SDSS-III and -IV. DR13 includes all SEQUELS data, giving the largest SDSS sample to date of spectra targeting intermediate redshift ranges.  SDSS-IV also allocated a significant number of fibers on the eBOSS plates to two additional dark-time programs. TDSS\citep[{\it T}ime {\it D}omain {\it S}pectroscopic {\it S}urvey;][]{morganson2015} seeks to understand the nature of celestial variables by deliberately targeting  objects that vary in combined SDSS DR9 and Pan-STARRS1 data \citep[PS1;][]{2002SPIE.4836..154K} . A large number of the likely quasar targets so selected are also targeted by the main eBOSS algorithms and therefore meet the goals of both surveys. TDSS-only targets fill $\sim$ 10 spectra per square degree. The main goal of the SPIDERS ({\bf Sp}ectroscopic {\bf Id}entification of  {\bf eR}OSITA {\bf S}ources) survey is to characterize a subset of X-ray sources identified by eROSITA \citep[{\bf e}xtended {\bf Ro}entgen {\bf S}urvey with an {\bf I}maging {\bf T}elescope {\bf A}rray;][]{2014SPIE.9144E..1TP} . Until the first catalog of eROSITA sources is available, SPIDERS will target sources from the RASS \citep[{\bf Ro}etgen {\bf A}ll {\bf S}ky {\bf S}urvey;][]{RASS} and XMM \citep[{\bf X}-ray {\bf M}ulti-mirror {\bf M}ission;][]{XMM}. SPIDERS will also obtain on average $\sim$ 10 spectra per square degree over the course of SDSS-IV, but the number of fibers per square degree on a plate is weighted toward the later years to take advantage of the new data from eROSITA. 
 
In bright time at APO, APOGEE-2, the successor to APOGEE (hereafter APOGEE-1) in SDSS-IV, will continue its survey of the Milky Way stellar populations. Critical areas of the Galaxy, however, cannot be observed from APO, including the more distant half of the Galactic bar, the fourth quadrant of the disk, and important dwarf satellites of the Milky Way, such as the Magellanic Clouds and some dwarf spheroidals. SDSS-IV will for the first time include operations outside of APO as the result of Carnegie Observatories and the Chilean Participation Group joining the collaboration. A second APOGEE spectrograph is being constructed for installation on the 2.5-meter du Pont Telescope \citep{1973ApOpt..12.1430B} at Las Campanas Observatory (LCO) near La Serena, Chile. When APOGEE-2S begins survey operations in 2017, approximately 300 survey nights on the du Pont Telescope will be used to extend the APOGEE-2 survey to the Southern Hemisphere.  

Data Release 13 (DR13) is the first data release for SDSS-IV, which will have regular public, documented data releases, in keeping with the philosophy of SDSS since its inception. In this paper, we describe the data available in DR13, focusing on the data appearing here for the first time. We present overall descriptions of the sample sizes and targeting and provide a detailed bibliography of the technical papers available to understand the data and the surveys in more detail. These technical papers and the SDSS website (\url{www.sdss.org}) contain critical information about these data, which here is only summarized.

\section{Overview of the Survey Landscape}
The release of DR13 coincides with the arrival of an astonishgly rich set of data from ongoing and recently
completed surveys outside of SDSS. \citet{overview}, \citet{Bundy_2015}, and \citet{Majewski_2017} describe how
the SDSS-IV surveys compare in survey strategy, size, and data within the wider arena of spectroscopic surveys. We cite here
some key science results for these works to complement our brief history of SDSS.

Spectroscopic redshift surveys of large scale structure have resulted in BAO
measurements over a range of redshifts. The BAO signal has been detected at lower redshift (z$\sim$0.1) from measurements
of $\sim$75,000 galaxies in the 6dF Galaxy Redshift Survey \citep{2011MNRAS.416.3017B}. WiggleZ measured the BAO signal at
similar redshifts to BOSS based on redshifts of $\sim$225,000 galaxies. The final WiggleZ results for the 1-D BAO peak 
\citep{2014MNRAS.441.3524K} and the 2-D BAO peak \citep{2017MNRAS.464.4807H} at z=0.44, 0.6 and 0.73 agree with the BOSS results. Still underway is the VIMOS
Public Extragalactic Redshift Survey, which focuses on higher redshifts than previous work ($0.5<$z$<1.2$) and will overlap
in part with eBOSS \citep{2014A&A...566A.108G}.

When the MaNGA survey began, two pioneering and highly influential IFU surveys of hundreds of galaxies were being
completed: CALIFA \citep{2012A&A...538A...8S} and ATLAS$^{\rm 3D}$ \citep{2011MNRAS.413..813C}. ATLAS$^{\rm 3D}$ observed 260
morphologically classified early-type galaxies (ETGs) within 40 Mpc at optical, radio, and infrared wavelengths, including
optical IFU observations with the PPAK integral field unit instrument. CALIFA released IFU data from SAURON on 667 galaxies in its third and final data
release \citep{2016A&A...594A..36S}. These galaxies spanned a range of morophologies from irregular to elliptical, over 7 magnitudes in
luminosity, and over 3 magnitudes in $u-z$ colors. The increase in the number of galaxies with IFU data led to many pivotal
discoveries regarding galaxy evolution, including advances in our knowledge of the origin and heating sources
of gas in ETGs, gas and stellar abundance gradients in galaxies, and the star-formation rate and age of stellar populations,
highlighted below.

\citet{2011MNRAS.417..882D} used ATLAS$^{\rm 3D}$ to investigate the origins of the gas in slow and fast rotating ETGs in
different environments. Overall, ETGs are poor in atomic and molecular gas and therefore lack much star formation,
and the cause of this transformation is critical to understanding galaxy formation. \citet{2011ApJ...735...88A} identified
one mechanism for gas depletion in the ATLAS$^{\rm 3D}$ galaxy NGC~1266, in the form of a strong molecular wind from
the nucleus, likely powered by a hidden AGN. \citet{2012A&A...540A..11K} suggested an AGN as an energy source
for the elongated ionized gas observed by CALIFA in NGC~5966. The CALIFA data also showed that NGC~5966 and its fellow
ETG, NGC~6762
had extended line emission best explained by heating from post-asymptotic giant branch (p-AGB) stars
\citet{2016A&A...588A..68G} studied 32 ETGs in CALIFA and found extended H$\alpha$ emission that fell into two broad classes
based on whether the intensity of emission is flat or increasing with radius.

The ages of stellar populations across CALIFA's wide range of Hubble types show that galaxies with stellar masses
$ > \sim 5 \times 10^9$ M$_{\odot}$ form from the inside out, with the more massive galaxies having older stellar
populations in each scaled radial bin \citep{2013ApJ...764L...1P}. \citet{2016ApJ...821L..26C} showed the relationship between star formation rate (SFR) \
and stellar mass that held for entire
star-forming galaxies was also true for spatially resolved regions within a galaxy. In another intriguing clue to the relation
between small and large scales, \citet{2014A&A...563A..49S} found that the HII regions in 306 CALIFA galaxies have a characteristic disk oxygen
abundance gradient when scaled to its effective radius.

Another large IFU survey, the SAMI survey \citep{2015MNRAS.447.2857B}, is currently underway at the Anglo-Australian Telescope
and will utimately observe 3400 galaxies. The Early Data Release \citep{2015MNRAS.446.1567A} provided data for 107 galaxies, while
the first major data release \citep{2017arXiv170708402G} includes 772 galaxies. Science
results from the SAMI survey so far include characterizing the galactic winds or extended diffuse ionizing gas in
edge-on disk galaxies \citet{2016MNRAS.457.1257H}, mapping the quenching of star formation proceeding from the outside-in in
dense environments \citet{2017MNRAS.464..121S}, and identifying stellar mass as the main variable affecting fraction of ETGs that
are slow rotators \citet{2017arXiv170401169B}

While APOGEE-2 is currently the only {\it infrared} stellar spectroscopic survey, there are several Galactic stellar
surveys of similar scope observing at optical wavelengths. The RAVE survey \citep{steinmetz2006} completed observations
in 2013 with R$\sim$7,500 spectra of $\sim$450,000 bright stars in the CaII triplet region released in Data Release 5
\citep{2017AJ....153...75K}. RAVE's primary goal was obtain spectroscopic measurements for stars with exquisite Gaia proper motion and
parallax measurements. The $\sim$ 250,000 stars in DR5 that
have proper motions and parallaxes in the first Gaia data release \citep{2016A&A...595A...2G} now make this work
possible in both the disk \citep[e.g.,][]{2017arXiv170406274R} and halo \citep[e.g.,][]{2017A&A...598A..58H}. Among the many results published
prior to the Gaia results were measurements of the local escape velocity \citep[e.g.,]{2007MNRAS.379..755S,2014A&A...562A..91P},
detection of a ``wobbly'' Galaxy from asymmetric velocities both radially and across the disk \citep{2011MNRAS.412.2026S,2013MNRAS.436..101W}), and the observation of extra-tidal stars from Galactic globular clusters \citep[e.g.,][]{2014A&A...572A..30K,2015A&A...583A..76F,2016MNRAS.457.2078A}

The Gaia-ESO survey recently completed observations of $\sim 10^5$ cluster and field stars at either R$\sim$20,000 or
R$\sim$47,000 \citep{2012Msngr.147...25G}.  The Gaia-ESO collaboration has presented important results about the nature
of several Galactic components, including the abundances and kinemetics in the bulge \citep[e.g.,][]{2014MNRAS.445.4241H,2017A&A...601A.140R,2017A&A...602L..14R}, characterizing the accreted component of the halo and (if any) disk \citep[e.g.,][]{2015MNRAS.450.2874R}, and measuring radial abundance gradients in the open cluster population \citep[e.g.,][]{2016A&A...591A..37J,2017A&A...603A...2M}.

The LAMOST Galactic survey is obtaining spectra, 4000 at a time, at a similar resolution to SEGUE \citep{zhao2012}.
Data Release 3\footnotemark[3]\footnotetext[3]{\url{dr3.lamost.org}} includes 5.75 million spectra and stellar parameters for $> 3$ million stars.
LAMOST has an ideal view of the field observed for four years by the {\it Kepler} satellite, and several groups have
used the $\sim$50,000 stars observed in the {\it Kepler} field to characterize its stellar populations 
\citep[e.g.,][]{2014ApJ...789L...3D,2016A&A...594A..39F,2016ApJS..225...28R,2017ApJ...834...92C,2017arXiv170607807D}. Other work based on LAMOST spectra includes 
measurements of stellar activity \citep[e.g.,][]{2016MNRAS.463.2494F} and identification of
important subclasses of objects from ultra metal-poor stars \citep[e.g.,][]{2015ApJ...798..110L} and metallitic lined Am stars \citep[e.g.,][]{2015MNRAS.449.1401H}.

The GALAH survey \citep{2015MNRAS.449.2604D}, which started operations in 2014, released spectra for over 200,000 stars
observed with the HERMES spectrograph (R$\sim$28,000) at the Anglo-Australian Telescope \citep{2017MNRAS.465.3203M}. When
the survey is completed, spectra of $\sim 10^6$ stars in 4 optical windows accessing 29 elements are expected.

\section{Scope of Data Release 13}

\begin{figure*}[h!]
\begin{center}
\includegraphics[width=5in]{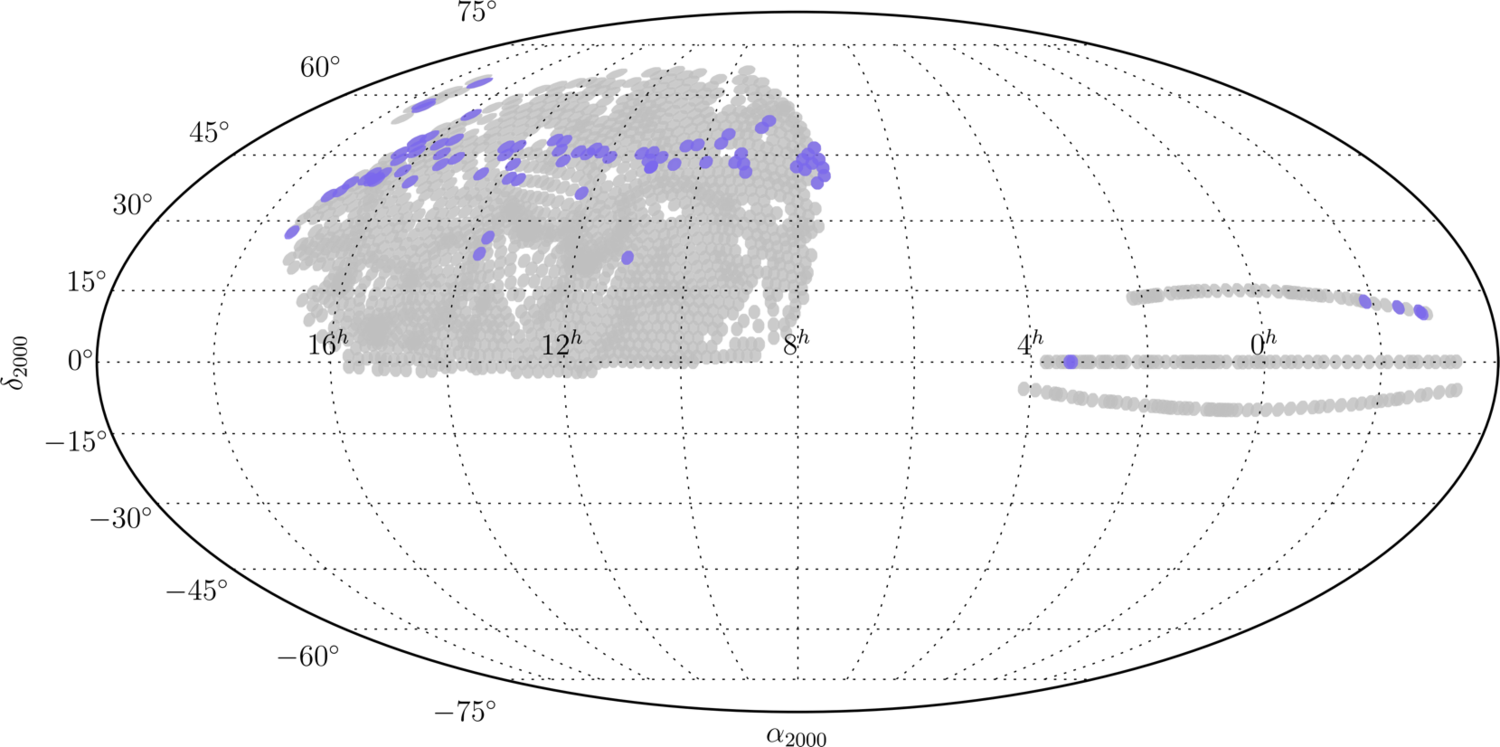}
\caption{{\label{manga_dr13}
In grey are shown the locations in equatorial coordinates of all possible plates with MaNGA galaxies, each containing 17 galaxy targets. Because the MaNGA targets are selected to have SDSS photometry, this footprint corresponds to the Data Release 7 imaging data \citep{2009ApJS..182..543A}. Approximately 30\% of these will be observed in the full planned MaNGA survey. The blue shows the plates observed in the first year of MaNGA for which data cubes are released in this paper.
}}
\end{center}
\end{figure*}

\begin{figure*}[h!]
\begin{center}
\includegraphics[width=5in]{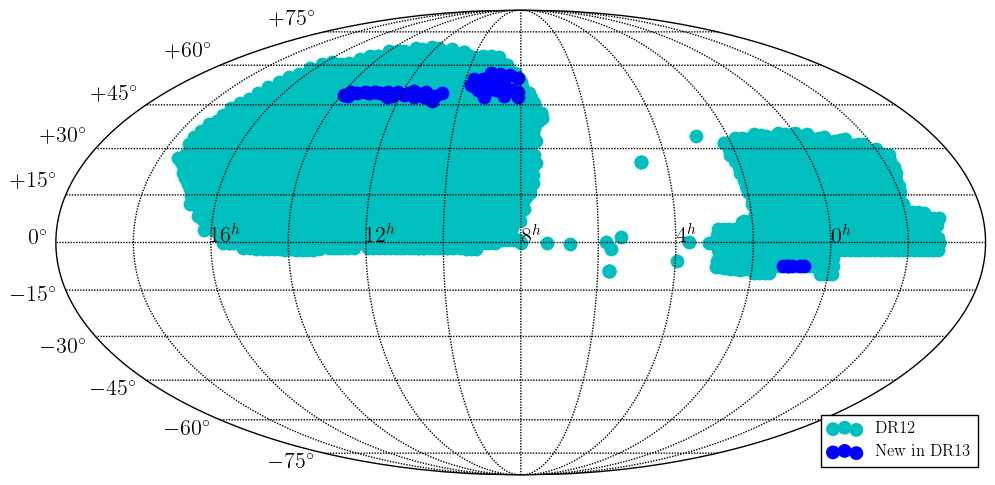}
\caption{{\label{eboss_dr13}
 Coverage of DR13 data from BOSS and SEQUELS in equatorial coordinates. The blue areas show the locations in equatorial coordinates of the five new BOSS plates (SGC) and the 126 SEQUELS plates (NGC) released in DR13. The green represents the area covered by BOSS in DR12. The SEQUELS plates released in DR12 lie in the same region as the new ones in DR13, providing complete coverage over roughly 400 square degrees. }}
\end{center}
\end{figure*}

\begin{figure*}[h!]
\begin{center}
\includegraphics[width=5in]{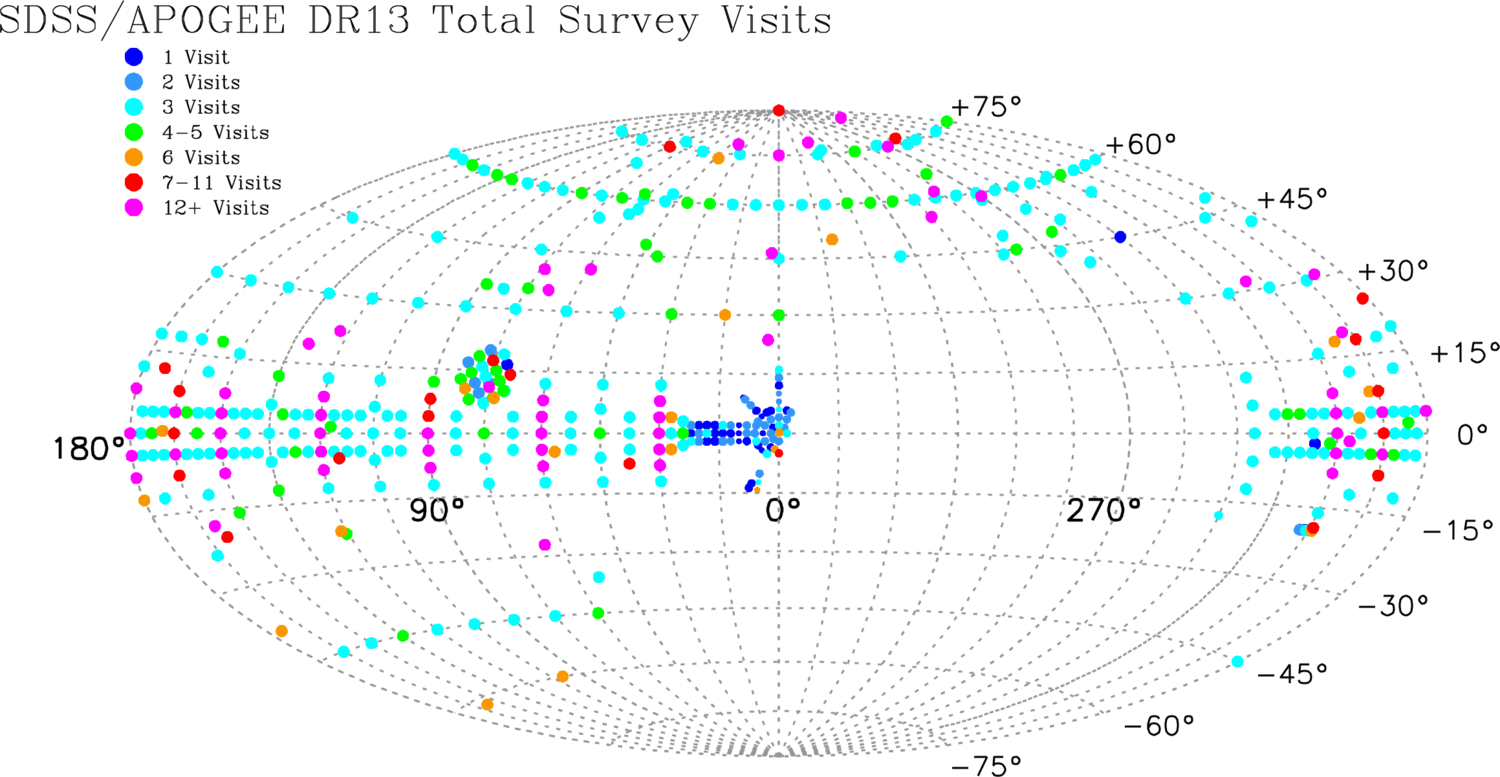}
\caption{{\label{apogee_dr13} Coverage of APOGEE-2 DR13 data in
    Galactic coordinates; the raw data and its coverage is the same as
    in DR12, but it has been reprocessed through the latest reduction
    pipeline and ASPCAP versions. The color coding denotes the number
    of visits to each field, as indicated in the legend.}}
\end{center}
\end{figure*}

SDSS-IV has been operating since July 2014. DR13 contains
data gathered between July 2014 and July 2015 and is summarized in
Table~1. The categories under MaNGA galaxies are
described in \S5. The SEQUELS targeting flags are listed and described
in \citet{2015ApJS..219...12A}. Figures~\ref{manga_dr13},~\ref{eboss_dr13},~\ref{apogee_dr13}
show the sky coverage of the MaNGA, eBOSS and APOGEE-2 surveys
respectively. In the subsequent sections, we discuss each survey's
data in detail, but briefly DR13 includes

\begin{itemize}

\item Reduced data for the 82 MaNGA galaxy survey plates, yielding
  1390 reconstructed 3-D data cubes for 1351 unique galaxies, that were completed by July 2015. Row-stacked
  spectra (RSS) and raw data are also included.
\item Reduced BOSS spectrograph data for an additional 60 SEQUELS
  plates, completing the SEQUELS program. The total number of SEQUELS
  plates released in DR12 and DR13 is 126. These plates provide a
  complete footprint covering roughly 400 square degrees that will not
  be revisited in eBOSS.  The targets include a superset of the eBOSS
  LRG and quasar samples, a sample of variability-selected point
  sources at a much higher density than in TDSS, and new
  X-ray-selected objects selected by similar criteria to targets in
  SPIDERS.
\item The reduced data for five BOSS plates at low declination in the
  SGC.  These plates were drilled during SDSS-III but not observed due
  to insufficient open-dome time when they were observable.  The
  plates were observed early in SDSS-IV to fill in the footprint in
  this region.
\item Spectroscopic data from BOSS processed with a new version of the data reduction pipeline, which results in less-biased flux values.
\item All APOGEE-1 data re-reduced with improved telluric subtraction and analyzed with an improved pipeline and synthetic grid, including adding rotational broadening as a parameter for dwarf spectra.
\item  New species (\ion{C}{1}, P, \ion{Ti}{2}, Co, Cu, Ge, and Rb) with reported abundances for APOGEE-1 sample. 
\item Stellar parameters for APOGEE-1 stars with cooler effective temperatures (T$_{\rm eff} < 3500$K), derived by an extension of the grid of synthetic spectra using MARCS \citep{marcs} model atmospheres.
\item Recalibrated SDSS imaging catalogs, using the hypercalibration to PanSTARRS-1 implemented by \citet{fink2016}.
\item Valued-added catalogs, see Table~2. More detail and direct links to the catalogs and their datamodels can be found at \url{http:\\www.sdss.org/dr13/data\_access/vac}.
\item The most recent reductions of all data from previous iterations of SDSS  is included as a matter of course. For MARVELS data, these data are the same as in DR12; for SEGUE and SEGUE-2, the same as in DR9.
\end{itemize}

\begin{deluxetable*}{lrr}
\label{bigtable}
\tablewidth{4in}
\tablecaption{Reduced spectroscopic data in DR13} 
\tablehead{ 
\colhead{Target Category} & \colhead{\# DR12} & \colhead{\# DR12+13} }
\startdata
\multicolumn{3}{l}{MaNGA main galaxy sample: } \\ 
\multicolumn{1}{r}{\tt PRIMARY\_v1\_2} & 0 & 600  \\ 
\multicolumn{1}{r}{\tt SECONDARY\_v1\_2} & 0  & 473  \\ 
\multicolumn{1}{r}{\tt COLOR-ENHANCED\_v1\_2} & 0 & 216  \\  
MaNGA ancillary galaxies\tablenotemark{1} & 0  & 31\\ 
MaNGA Other & 0 & 62  \\
\tableline 
\multicolumn{3}{l}{SEQUELS  } \T \\ 
\multicolumn{1}{r}{\tt LRG\_IZW}	 & 11781& 21271 \\	
\multicolumn{1}{r}{\tt LRG\_RIW} & 11691 & 20967 \\
\multicolumn{1}{r}{\tt QSO\_EBOSS\_CORE} & 19455 & 33367 \\	
\multicolumn{1}{r}{\tt QSO\_PTF} & 13227 & 22609	\\
\multicolumn{1}{r}{\tt QSO\_REOBS} & 1357 & 2238	\\
\multicolumn{1}{r}{\tt QSO\_EBOSS\_KDE}	& 11836 & 20474\T \\	
\multicolumn{1}{r}{\tt QSO\_EBOSS\_FIRST} & 293 &519	\\
\multicolumn{1}{r}{\tt QSO\_BAD\_BOSS}	& 59 & 95	\\
\multicolumn{1}{r}{\tt QSO\_BOSS\_TARGET} &	59 & 95	\\	
\multicolumn{1}{r}{\tt DR9\_CALIB\_TARGET} &	28594 & 49765 \\	
\multicolumn{1}{r}{\tt SPIDERS\_RASS\_AGN}	&162	 & 275\T \\	
\multicolumn{1}{r}{\tt SPIDERS\_RASS\_CLUS}	 &1533  & 2844 \\	
\multicolumn{1}{r}{\tt TDSS\_A}	& 9412	 & 17394 \\	
\multicolumn{1}{r}{\tt TDSS\_FES\_DE} & 40 &  70	\\
\multicolumn{1}{r}{\tt TDSS\_FES\_DWARFC} &19 & 29 \\	
\multicolumn{1}{r}{\tt TDSS\_FES\_NQHISN} & 73 & 148\T \\	
\multicolumn{1}{r}{\tt TDSS\_FES\_MGII} &1 & 2	\\
\multicolumn{1}{r}{\tt TDSS\_FES\_VARBAL} & 55 & 103 \\	
\multicolumn{1}{r}{\tt SEQUELS\_PTF\_VARIABLE} & 700 &1153	\\
\tableline 
\multicolumn{3}{l}{APOGEE-2} \T \\
\multicolumn{1}{r}{All Stars} & 164562 & 164562 \\
\multicolumn{1}{r}{NMSU 1-meter stars} & 894 & 894 \\
\multicolumn{1}{r}{Telluric stars} & 17293 & 17293 \\
\multicolumn{1}{r}{Commissioning stars} & 11917 & 11917 \\
\tablenotetext{1}{Many MaNGA ancillary targets were also targeted as
  part of the main galaxy sample, and are counted twice in this table.} 
\enddata
\end{deluxetable*}

\begin{deluxetable*}{lll}
\tabletypesize{\scriptsize}
\label{vactable}
\tablecaption{Value-Added Catalogs New in DR13} 
\tablehead{ 
\colhead{Catalog Description} & \colhead{Reference} & \colhead{\url{http://data.sdss.org/sas/dr13/}}}
\startdata
SPIDERS Clusters demonstration sample catalog & \citet{2016MNRAS.463.4490C} & \url{eboss/spiders/analysis/} \\
SPIDERS AGN target selection catalog & \cite{2017arXiv170401796D} & \url{eboss/spiders/target/} \\
SPIDERS cluster target selection catalog & \citet{2016MNRAS.463.4490C}  & \url{eboss/spiders/target/} \\
WISE Forced Photometry & \citet{2016AJ....151...36L} & \url{eboss/photoObj/external/WISEForcedTarget/301/}\\
Composite Spectra of Emission-line Galaxies & \citet{2015ApJ...815...48Z} &  \url{eboss/elg/composite/v1\_0/} \\
ELG Fisher selection catalog & \citet{delubac2017} & \url{eboss/target/elg/fisher-selection/} \\
Redmonster redshift \& spectral classification catalog & \citet{2016AJ....152..205H} & \url{eboss/spectro/redux/redmonster/v5\_9\_0/v1\_0\_1/}\\
QSO Variability & \citet{qsovar} & \url{eboss/qso/variability/} \\
APOGEE DR13 red-clump catalog & \citet{bovy2014} &  \url{apogee/vac/apogee-rc/cat/}\\
\enddata
\end{deluxetable*}

DR13 contains only a subset of the reduced or raw data for all surveys taken between July 2014 and July 2015. The first {\it two} years of eBOSS data are needed before useful cosmological constraints can be extracted. APOGEE-2 is using the first year of SDSS-IV data to work on science verification and targeting optimization for new classes of targets and new survey strategies. Both of these surveys will release more extensive new data in Data Release 14.
\section{Data Distribution}
The data for DR13 are distributed through the same mechanisms available in DR12, with some significant changes to the environment used to access the catalog data (see below). Raw and processed image and spectroscopic data are available, as before, through the Science Archive Server (SAS, \url{data.sdss.org/sas/dr13}), and also for imaging data, optical spectra, and APOGEE IR spectra through an interactive web application (\url{dr13.sdss.org}, available soon). The catalogs of photometric, spectroscopic, and derived quantities are available through the Catalog Archive Server or CAS \citep{2008CSE....10...30T,2008CSE....10....9T} via two primary modes of access: browser-based queries of the database are available through the SkyServer Web application (\url{http://skyserver.sdss.org}) in synchronous mode, and more advanced and extensive querying capabilities are available through CasJobs (\url{http://skyserver.sdss.org/casjobs}) in asynchronous or batch mode that allows time-consuming queries to be run in the background \citep{2008CSE....10...18L}.

The CAS is now part of the new SciServer (http://www.sciserver.org/) collaborative science framework that allows users single-sign-on access to a suite of collaborative data-driven science services that includes the classic SDSS services of SkyServer and CasJobs. These services are essentially unchanged in their user interfaces but have acquired powerful new capabilities and undergone fundamental re-engineering to make them interoperable and applicable to other science domains. New services are also available to users once they register on the SciServer portal, and these services work seamlessly with the existing tools. Most notable among the new offerings are SciDrive, a distributed DropBox-like file store; SkyQuery, a federated cross-matching service that compares and combines data from a collection (federation) of multi-wavelength archives (SkyNodes); and SciServer Compute, a powerful new system for uploading complex analysis scripts as Jupyter notebooks (using Python, MatLab or R) running in Docker containers.

Links to all of these methods are provided at the SDSS website (\url{http://www.sdss.org/dr13/data\_access}) The data processing software for APOGEE, BOSS, and SEGUE are publicly available at \url{http://www.sdss.org/dr13/software/products}. A set of tutorial examples for accessing SDSS data is provided at \url{http://www.sdss.org/dr13/tutorials}. All flat files are available for download at \url{http://data.sdss.org/datamodel/}

\section{Recalibration of Imaging Data}

DR13 includes a photometric recalibration of the SDSS imaging data. \citet{fink2016} rederived the $g$, $r$, $i$ and $z$ band zero points using the PS1 photometric calibrations of 
\citet{2012ApJ...756..158S}, as well as rederiving the flat fields in all five bands (including $u$) .
This effort improved the accuracy of the SDSS photometry previously hindered
by a paucity of overlapping scans to perform u\"bercalibration across the entire SDSS sky and by sub-optimal flatfields. 
The residual systematics, as measured by comparison with PS1 photometry and spectral energy
distributions for stars with spectra, are reduced to 0.9, 0.7, 0.7 and 0.8\% in the $griz$ bands, 
respectively; several previously uncertain calibrations of specific runs are also now much better constrained. The 
resulting recalibrated imaging catalogs are the basis for the eBOSS and MaNGA targeting.  

For the MaNGA target selection, we are using the NASA-Sloan Atlas (NSA; \citealt{blanton11}), a reanalysis of the SDSS photometric data using sky subtraction and deblending better tuned for large galaxies. Relative to the originally distributed version of that catalog, we have used the new calibrations mentioned above, increased the redshift range up to $z=0.15$, and have added an elliptical aperture Petrosian measurement of flux, which MaNGA targeting is based upon. DR13 includes the NSA catalog (version {\tt v1\_0\_1}) associated with this reanalysis as the {\tt nsatlas} CAS table and as a file on the SAS. For the MaNGA galaxies released in  DR13, we provide the actual images (referred to in MaNGA documentation as ``preimaging'')  on the SAS as well.

\citet{2016AJ....151...36L} reanalyzed data from the {\bf W}ide-field {\bf I}nfrared {\bf S}atellite {\bf E}xplorer (WISE; \citealt{wright10a}) to use for eBOSS targeting. They used positions and galaxy profile measurements from SDSS photometry as input structural models and constrained fluxes in the WISE 3.4 $\mu$m and  4.6 $\mu$m bands. These results agree with the standard WISE reductions to within 0.03 mag for high signal-to-noise ratio, isolated point sources in WISE. However, the new reductions provide flux measurements for low signal-to-noise ratio  ($<5\sigma$) objects detected in the SDSS but not in WISE (over 200 million objects). Despite the fact that the objects are undetected, their flux measurements are nevertheless informative for target selection, in particular for distinguishing stars from quasars. This photometry is provided as a value-added catalog in the {\tt wiseForcedTarget} CAS table and on the SAS as described in Table 2. 

The Galactic extinction estimates published in the SDSS imaging tables ({\tt photoObj} and related tables in the CAS) have been changed. The Galactic extinction still uses the \citet{1998ApJ...500..525S} models of dust absorption to estimate $E(B-V)$,  but the Galactic extinction coefficients for each band have been updated as recommended by  \citet{2011ApJ...737..103S}.  The extinction coefficients $R_u$, $R_g$, $R_r$, $R_i$, and $R_z$ are changed from the values used in BOSS (5.155, 3.793, 2.751, 2.086, 1.479) to updated values (4.239, 3.303, 2.285, 1.698, 1.263). The corresponding numbers for the WISE bands are $R_{W1}=0.184$ for the WISE 3.4$\mu$m band and $R_{W2}=0.113$ for the 4.6$\mu$m band \citep{1999PASP..111...63F}.
\section{MaNGA: Integral Field Spectroscopic Data }

MaNGA is investigating the internal kinematics and composition of gas
and stars in low redshift ($z\leq0.15$) galaxies using fiber bundles to feed the BOSS
spectrographs. \citet{Bundy_2015} describe the high-level scientific
goals, scope, and context of the survey in investigating galaxy
formation while \citet{yan2016b} give a detailed description of the
survey design, execution, and data quality relevant to DR13.  With
nearly 1390 data cubes released, MaNGA's DR13 data products represent
the largest public sample to date of galaxies
observed with integral field spectroscopy. This data set signifies a
valuable first step in MaNGA's goals to reveal the internal properties
and dynamics of a statistically powerful sample of galaxies, that spans a broad
range in stellar mass, local environment, morphology, and star
formation history. Individual observations across the sample are of
sufficient quality to characterize the spatially-dependent composition
of stars and gas as well as their internal kinematics, thus providing
important clues on growth and star formation fueling, the build-up of
spheroidal components, the cessation of star formation, and the
intertwined assembly history of galaxy subcomponents.

The survey was made possible through an instrumentation initiative in
SDSS-IV to develop a reliable and efficient way of bundling 1423
optical fibers into tightly-packed arrays that constitute MaNGA's IFUs \citep{Drory_2015}. 
For each pointing, MaNGA observes 17 science targets with IFUs ranging from 19
to 127 fibers (with diameters of 12$-$32$\arcsec$). The IFU size
distribution was optimized in concert with the sample design \citep{Wake_2017} that targets SDSS-I/II main sample galaxies at a
median redshift of 0.03 to obtain in 6 years a sample of
10$^4$ galaxies with uniform radial coverage and a roughly flat
distribution in $\log {\rm M}_*$ limited to M$_* > 10^9{\rm M}_{\odot}$. Careful
attention was paid to optimizing hardware and an observing strategy
that ensures high quality imaging spectroscopy \citep{Law_2015} and to
surface photometric flux calibration with a precision better than 5\%
across most of the wavelength range, 3,622-10,354\AA{}\citep{yan2016a}. 
As described in these papers, salient aspects
included protocols for constraining hour-angles of observations
to limit differential atmospheric diffraction, dithering exposures to
avoid gaps in the coverage of the targets because of space between the 
fibers, and special calibration mini-bundles to ensure reliable absolute and relative flux
calibration. An automated pipeline delivers sky-subtracted,
flux-calibrated row-stacked-spectra (RSS) and datacubes for all
sources \citep{law2016}.

\subsection{MaNGA DR13 Main Galaxy Sample}

At the completion of SDSS-IV, the MaNGA survey's main galaxy sample
will include $\sim10^4$ galaxies with ${\rm M_*} > 10^9 {\rm M}_\odot$
and a roughly flat stellar mass distribution. DR13's 1390 galaxy
data cubes, corresponding to 1351 unique galaxies makes it the largest public sample of galaxies with IFU
spectroscopy. MaNGA's main galaxy sample consists of three major
parts: Primary sample, Secondary sample, and the Color-Enhanced
supplement. 

\begin{figure*}[htb]
\begin{center}
\includegraphics[width=5in]{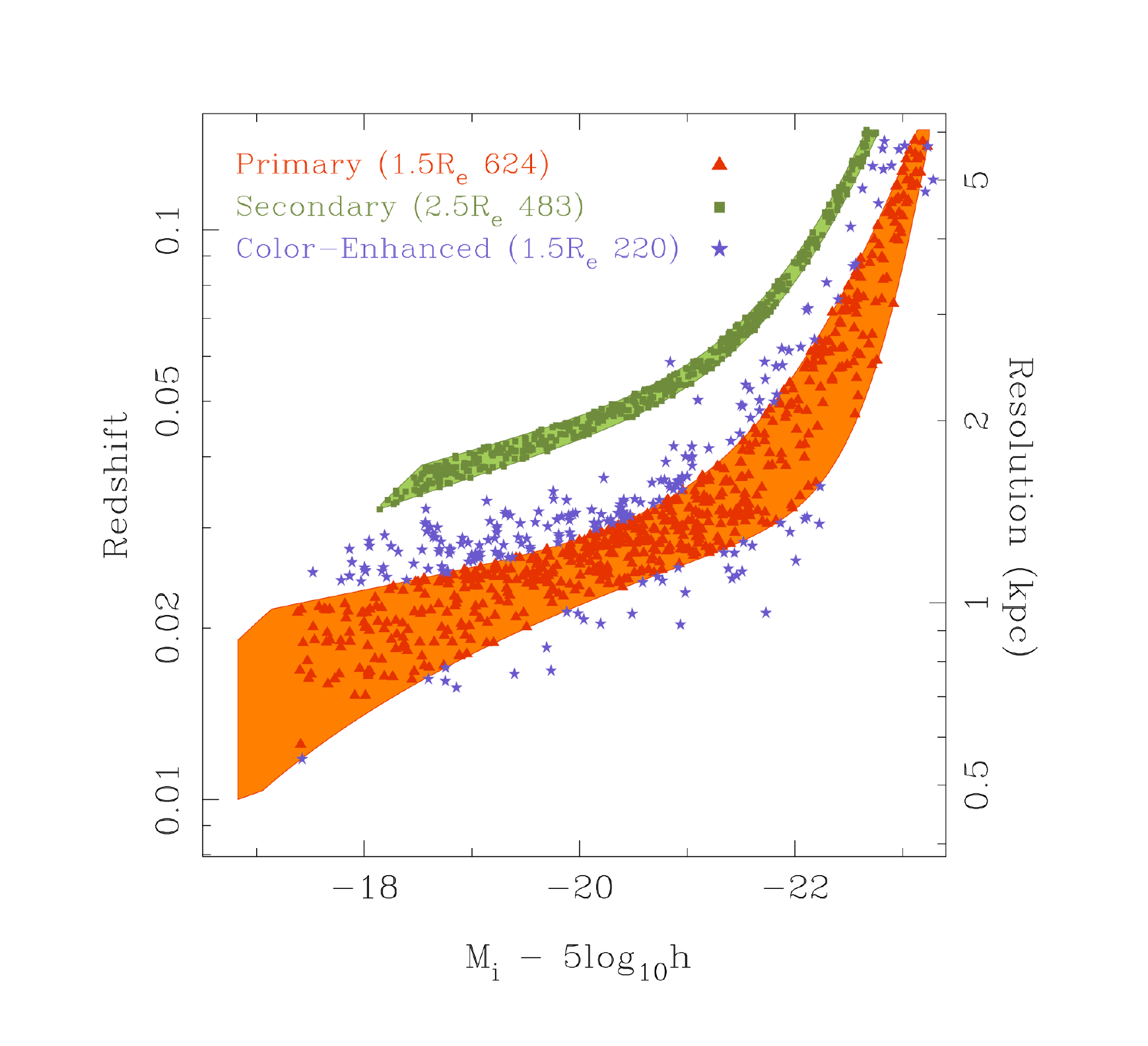}
\caption{{Principal selection cuts for the main MaNGA samples, where
    $h = H_0 / 100 \ {\rm km \ s}^{-1}$. The colored bands
    show the selection cuts for the Primary (orange) and Secondary
    (green) samples illustrating the M$_i$ dependence of the redshift
    limits. More luminous, and hence typically larger galaxies are
    selected at higher redshift than less luminous galaxies, ensuring
    that the angular size (1.5 $R_e$ or 2.5 $R_e$) distribution is
    roughly independent of luminosity. The volume sampled also
    increases as the luminosity increases in such a way as to ensure a
    constant number density of galaxies at all luminosities. The
    points show the positions in this plane for the MaNGA galaxies in
    DR13, for the Primary (red triangles), Secondary (green squares)
    and Color-Enhanced samples (blue stars), although the
    Color-Enhanced selection also depends on NUV-$i$color (see text for
    details). The numbers in the legend give the total number of observations
    of galaxies in each class, including repeat observations. The right hand y-axis gives an indication of the
    physical size of the mean spatial resolution element of the MaNGA
    data.}}
\label{fig:mangaselection}
\end{center}
\end{figure*}

\begin{figure*}[htb]
\begin{center}
\includegraphics[width=5in]{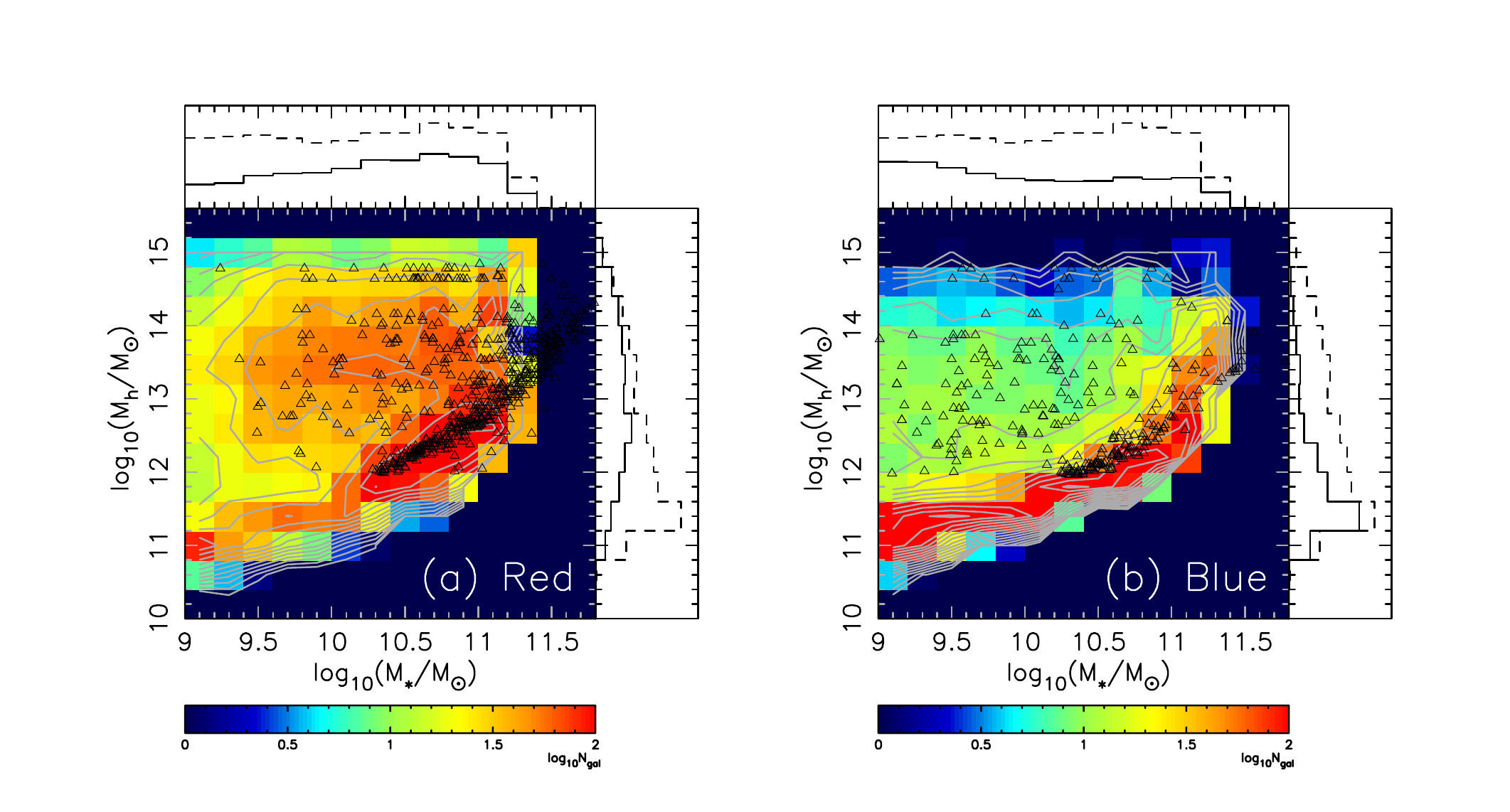}
\caption{{The location of MaNGA galaxies (black triangles) in this data release area in 
    the plane of stellar mass (M$_*$, $x$-axis)
    versus dark matter halo mass (M$_h$, $y$-axis). The two panels
    show red and blue galaxies separately,which are divided by a
    single color cut at $g-r=0.7$. Stellar masses are
    taken from the NASA-Sloan Atlas, and halo masses
    from SDSS/DR7, using the method of \citet{2007ApJ...671..153Y}. 
    Plotted as a colormap in the background are the number of MaNGA galaxies
    predicted for a 6-year full survey based  on mock catalogs
    informed by the semi-analytic model of \citet{2011MNRAS.413..101G} 
    and constructed as in \citet{2006MNRAS.368...21L}, which are the same as Figure 3 of
    \citet{Bundy_2015}. Normalized histograms show
    1D marginalized M$_*$ distributions (top axes) and M$_h$
    distributions (right axes), with  dashed lines for the full
    primary sample and solid lines for the red (left) and blue (right)
    populations.  }}
\label{mangamass}
\end{center}
\end{figure*}

The Primary sample and the Secondary sample are selected from two
luminosity-dependent redshift bands, as shown in
Figure~\ref{fig:mangaselection}. The selection for each sample is
volume-limited at each absolute $i$-band magnitude. The shape of the
redshift bands is motivated by MaNGA's science requirements of having a
uniform spatial {\it coverage} in units of galaxy's effective radius
$R_e$ and having a roughly flat stellar
mass distribution \citep{yan2016b,Wake_2017}. Figure~\ref{mangamass} shows
the distribution of the MaNGA DR13 galaxies in the stellar mass vs. dark matter halo
mass plane. Because more
massive galaxies are on average larger, we observe them at a larger
distance than low mass galaxies. We chose the redshift bands so that
the great majority of the Primary (Secondary) sample is covered by our
fiber bundles to 1.5 $R_e$ (2.5 $R_e$) along their major axes. This
has the commensurate effect of changing the physical resolution
systematically as a function of stellar mass, as illustrated in
Figure~\ref{fig:mangaselection}. Potential deleterious effects of this
change in sampling are addressed by an ancillary program, described
below.

We also designed a Color-Enhanced supplement, as a supplement to the
Primary sample, to enhance the sampling of galaxies with rare
color-magnitude combinations, such as low-mass red galaxies, high mass
blue galaxies, and green valley galaxies. This is achieved by
extending the redshift limits around the Primary sample redshift band
for each underpopulated region in color-magnitude space.

The combination of the Primary sample and the Color-Enhanced
supplement is referred to as the Primary+ sample. We provide in our
data release the redshift limits over which each galaxy is
selected. This permits a correction to the sample using $1/V_{\rm
  max}$ weight to reconstruct a volume-limited representation of the
galaxy population, provided that there is negligible galaxy
evolution over this limited redshift range. More details of how we arrived at this selection
can be found in \cite{yan2016b} and \citet{Wake_2017}.  \citet{Wake_2017} provides the details of how to properly weight the
sample to reconstruct a volume-limited representation.

Among the 1351 unique galaxies released in DR13, there are 600 Primary
Sample galaxies, 473 Secondary Sample galaxies, and 216 Color-Enhanced
supplement galaxies. There are 62 galaxies that do not belong to any
of the above. Some of these are ancillary program targets (see below),
some are filler objects on plates with spare bundles, and others are
galaxies selected using older, obsolete versions of the selection and
observed on early plates. For most statistical analyses, these 62
galaxies should be excluded.

Which sample a given target galaxy belongs to is given by the
MANGA\_TARGET1 bitmask (or mngtarg1 in the ``drpall'' file; see \S5.2). Primary sample
galaxies have bit 10 set to 1, Secondary Sample galaxies have bit 11
set to 1, and Color-Enhanced Supplement Galaxies have bit 12 set to
1. Bits 1-9 are for obsolete selections and should be ignored.

\subsubsection{MaNGA Galaxy Ancillary Programs}

Roughly 5\% of the IFUs are assigned to ancillary science programs
defined by and allocated through internal competition and review. 
This assignment takes advantage of sky regions with a low
density of galaxies defined by our main survey criteria (above) or
where certain rare classes of objects, possibly outside our selection
cuts, are of sufficiently high science value to re-allocate IFUs from
the main program. Such high-value targets sometimes come from the main
sample, but the ancillary science goals require a different bundle
size, a slightly different center positions, or higher prioritization
over the random selection among the main sample galaxies.  There are
also science cases where using observing strategies different from
standard MaNGA observations are required. These lead to dedicated
plates.

We solicited ancillary proposals in all these categories during the
first year of survey observations, and they started to be included in
plate design half way through this year. Consequently, the ancillary
fraction in DR13 is smaller than 5\%. Some ancillary programs have
tens of targets approved but only a few got observed during the first
year, while some have no observations during this period. More targets
for these programs will be observed in the future. To identify
ancillary targets, one should use the MANGA\_TARGET3 bitmask (or
mngtarg3 in the drpall file). All ancillary targets have MANGA\_TARGET3
greater than zero. Additional information on the scientific
justification and targeting for each ancillary program can be found at
\url{http://www.sdss.org/dr13/managa/manga-target-selection/ancillary-targets}. Here
we provide some highlights and the corresponding bitnames.

\begin{itemize}

\item \textit{Luminous AGN:} This program increases the number of host
  galaxies of the most luminous active galactic nuclei (AGN). The first source of
  targets is the BAT 70-month Hard X-ray catalog
  \citep{2013ApJS..207...19B}. These have the bitname
  {\tt AGN\_BAT}. To increase the sample size further we used the
         [OIII]-selected catalog of \cite{2013MNRAS.433..622M}
         (bitname {\tt AGN\_OIII}). To match the distribution of
         bolometric luminosities between the samples, we selected 5
         [OIII] objects at comparable L$_{\rm bol}$ to each BAT object,
         within a redshift range of 0.01-0.08. The bolometric
         corrections are from \citet{2013MNRAS.436.3451S} and
         \citet{2009MNRAS.392.1124V}.

\item \textit{Edge-On Starbursts:} We will use edge-on starbursts to study the morphology and ionization state of large-scale outflows. To identify good targets, the specific star formation rate (sSFR) and inclination of every object in the baseline MaNGA targeting catalog was calculated. The sSFR was determined using WISE photometry from \citet{2016AJ....151...36L} and the calibration between the W4 filter and 22 $\mu$m emission in \citet{2013AJ....145....6J} . We then use a calibration from \cite{2014ApJ...782...90C} to derive the sSFR. The axis ratio {\tt SERSIC\_BA} in the targeting catalog was used to derive the inclination. All targets in this program have log sSFR$> -8.75$ and inclination $> 75$ deg.. The four galaxies in DR13 that have these properties, but were not included in the main galaxy target sample, have bitname {\tt EDGE\_ON\_WINDS}.

\item \textit{Close Pairs and Mergers:} Interactions and mergers can
  play a key role in galaxy evolution, and therefore an ancillary
  program was accepted that either slightly adjusted the field centers
  for some targets already included in the main galaxy sample or
  placed new IFUs on galaxies. Close pairs are defined as galaxies in
  the NSA catalog or the SDSS group catalog of
  \citet{2007ApJ...671..153Y} and X. Yang (private communication)
  with projected separation $<$ 50 kpc $h^{-1}$ and line-of-sight
  velocity (dV) $<$ 500 km s$^{-1}$, if both redshifts are
  available. If the bitname is {\tt PAIR\_ENLARGE}, then to get the
  full pair required a larger IFU than the one originally scheduled by
  the targeting algorithm \citep{Wake_2017}. If the bitname is
  {\tt PAIR\_RECENTER} this means the original assigned MaNGA IFU is
  sufficiently large, but requires re-centering. In addition to these
  already-planned galaxies, two sources of new objects were used. The
  one galaxy in DR13 with {\tt PAIR\_SIM} comes from the Galaxy Zoo
  Mergers Sample \citep{2016MNRAS.459..720H}. A critical sample comes
  from the ancillary program that requests that each galaxy be assigned a 
  separate IFU {\tt PAIR\_2IFU}. Only four pairs of interacting galaxies are
  serendiptiously targeted in the main MaNGA galaxy sample with separate IFUs,
  and this sample will compensate for the strong bias in the single IFU sample towards close separations 
  or higher redshifts.

\item \textit{Massive Nearby Galaxies:} Because the largest MaNGA IFU
  covers 32$\arcsec$, more luminous, larger galaxies observed to the same
  effective radius have poorer spatial resolution. The one {\tt
    MASSIVE} ancillary target in DR13 is part of a program to obtain a
  sample of nearby large galaxies with spatial resolution better than
  3 kpc and similar to the faintest galaxies in the MaNGA primary
  sample, at the cost of spatial extent.

\item \textit{Milky Way Analogs:} \citet{2015ApJ...809...96L} defined a
  sample of Milky Way analogs based on M$_*$, SFR, absolute magnitudes,
  and colors. MaNGA is including some of these analogs in the main
  galaxy catalog, but they are slightly biased or deficient in certain
  regions of parameter space. Galaxies with the bitname {\tt MWA} are
  drawn from the \citet{2015ApJ...809...96L} catalog to provide
  galaxies in those under-represented parts of parameter space.

\item \textit{Dwarf Galaxies in MaNGA:} The MaNGA main galaxy sample
  has galaxies with M$_* > 10^9M_{\odot}$, but dwarf galaxies are the
  most numerous galaxies in the Universe. This ancillary program
  provides 2 dwarf galaxies with MaNGA observations in DR13, the first
  observations of a larger sample expected by the end of the survey
  covering a range of environments. These galaxies are indicated by
  the bitname {\tt DWARF} and are drawn from the
  \citet{2012ApJ...757...85G} galaxy catalog with stellar
  masses $<10^9M_{\odot}$.

\item \textit{Brightest Cluster Galaxies:} The brightest cluster galaxies (BGCs) targeted here are
  brighter and in more massive halos than BCGs already in the MaNGA
  main sample and have the bitname {\tt BCG}. We base our target selection on the updated
  \cite{2007ApJ...671..153Y} cluster catalog, created from the SDSS
  DR7 NYU VAGC, an update of the DR4 version of
  \citet{2005AJ....129.2562B}

\item \textit{MaNGA Resolved Stellar Populations:} The ancillary
  program targets NGC 4163, a nearby dwarf galaxy with existing HST
  imaging and high-quality color-magnitude diagrams selected from the
  ACS Nearby Galaxy Survey \citep{2009ApJS..183...67D}. This galaxy is
  flagged by the bitname {\tt ANGST}.
\end{itemize}

\begin{deluxetable*}{lrrrcr}
\tablecaption{Summary of MaNGA Ancillary Programs with Data in DR13}
\tablehead{
\colhead{Ancillary Program}  & \colhead{\# of Targets in DR13}\tablenotemark{1} & \colhead{\# of Total Targets}\tablenotemark{1,2} &\colhead{BITNAME} & \colhead{binary digit}  & \colhead{value} 
}
\startdata
Luminous AGN   & 1 & 267 &{\tt AGN\_BAT} & 1  & 2 \\ 
                           & 4 & & {\tt AGN\_OIII}  & 2 & 4 \\ 
  Edge-On Star-forming Galaxies        & 4 & 166 &  {\tt EDGE\_ON\_WINDS} & 6 & 64 \\ 
 Close Pairs and Mergers        & 5 & 510 &{\tt PAIR\_ENLARGE} & 7 & 128 \\ 
         & 10 & & {\tt PAIR\_RECENTER} & 8 & 256 \\ 
         & 1 & & {\tt PAIR\_SIM} & 9 & 512 \\ 
         & 1 & & {\tt PAIR\_2IFU} & 10 & 1024 \\ 
Massive Nearby Galaxies         & 1 & 310& {\tt MASSIVE} & 12 & 4096 \\ 
   Milky Way Analogs      & 2 & 250& {\tt MWA} & 13 & 8192 \\ 
    Dwarf Galaxies in MaNGA     & 2 & 247 &{\tt DWARF}  & 14 & 16384 \\ 
      Brightest Cluster Galaxies   & 2 & 378 &{\tt BCG} & 17 & 131072 \\ 
       MaNGA Resolved Stellar Populations  & 1 & 4 & {\tt ANGST} & 18 & 262144 \\
    \enddata 
    \tablenotetext{1}{An individual galaxy can be targeted by more than one ancillary program.}
    \tablenotetext{2}{Number for each Ancillary Program refers to all targets in that program, regardless
of bit name}
\end{deluxetable*}

\subsection{MaNGA Data Products: Individual Fiber Spectra and 3-D Data Cubes}

\begin{figure*}[htb]
\begin{center}
\includegraphics[width=5in]{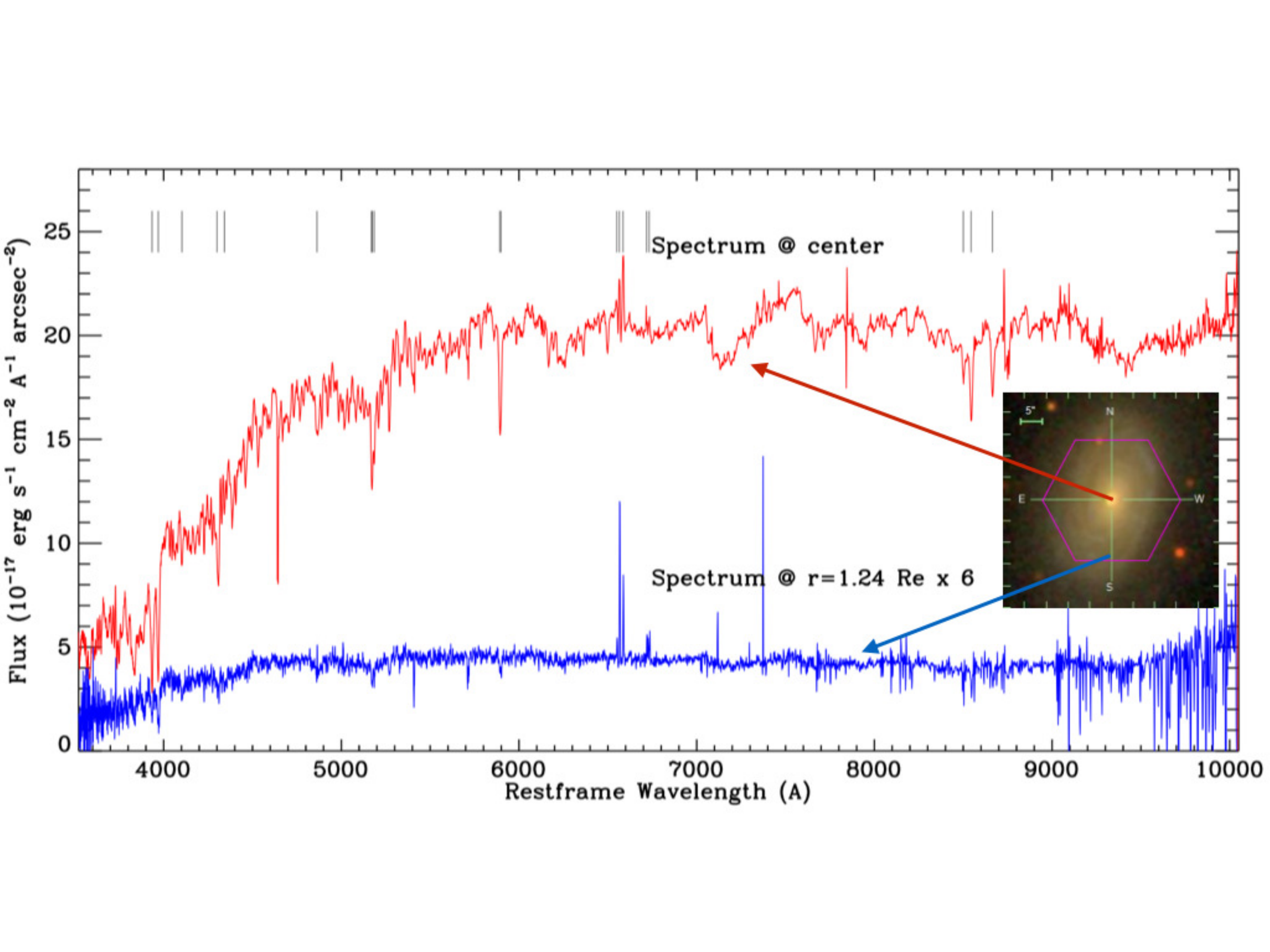}
\caption{{\label{mangaspec} Example spectra from a typical MaNGA data cube, adapted from \citet{yan2016b}. The inset shows the SDSS color image with the hexagonal IFU footprint overlayed. The top spectrum is from the central spaxel; the bottom spectrum from a spaxel 1.2 $R_e$ from the center and is multiplied by a factor of 6 for easier comparison with the central spectrum. The differences in the stellar and gas components between the two regions can clearly be seen, as well as the large number of diagnostic features to understand the star formation history and the physical conditions of the gas.
}}
\end{center}
\end{figure*}

In DR13, MaNGA is releasing both raw data (in the form of individual
CCD frames) and reduced data produced by version 1\_5\_4 of the MaNGA
Data Reduction Pipeline (DRP).  Figure~\ref{mangaspec} illustrates the quality of the
spectra from this pipeline. The MaNGA observing strategy is
described by \cite{Law_2015}, and the flux calibration by
\cite{yan2016a}.  Details of the MaNGA DRP, data products, and data
quality are given by \citet{law2016} (hereafter L16).  All MaNGA data are
in the form of multi-extension FITS files.

The DRP data products consist of intermediate reduced data
(sky-subtracted, flux-calibrated fiber spectra with red and blue data
combined for individual exposures of a plate) and final-stage data
products (summary row-stacked spectra and data cubes) for each target
galaxy.  The summary row stacked spectra (RSS files) are
two-dimensional arrays provided for each galaxy in which each row
corresponds to a single fiber spectrum. For a 127-fiber IFU with 9
exposures, there are thus 127 $\times$ 9 = 1143 rows in the RSS file.
These RSS files have additional extensions giving astrometric
information about the wavelength-dependent locations of each fiber on
the sky.

The three-dimensional data cubes (axes R.A., Decl., wavelength) are
created by combining the individual spectra for a given galaxy
together onto a regularized 0.5$\arcsec$ grid (see L16 for more detail).
Both data cubes and RSS files are provided in a version with a log
wavelength scale, which is the standard extraction and is relatively
smooth in velocity space, and in a version with a linear wavelength
scale, created directly from the native pixel solution rather than by
resampling the log-scaled spectra resampling. Each MaNGA data cube has
associated extensions corresponding to the estimated inverse variance
per pixel and a bad-pixel mask containing information about the
quality of a given pixel within the cube (depth of coverage, bad
values, presence of foreground stars, etc).  Additional extensions
provide information about the instrumental spectral resolution,
individual exposures that went into the composite data cube,
reconstructed $griz$ broadband images created from the IFU spectra, and
estimates of the $griz$ reconstructed point spread function.

The objects observed by MaNGA for which data cubes have been produced
are summarized in the ``drpall'' file, a FITS binary table with one
entry per object (including both galaxies and spectrophotometric
standard stars observed with 7-fiber IFUs to calibrate the MaNGA
data).  This drpall file lists the name, coordinates, targeting
information (e.g., redshift as given by the NASA Sloan Atlas),
reduction quality, and other quantities of interest to allow users to
identify galaxy targets of interest.  We note that MaNGA adopts two
naming schemes.  The first, termed ``mangaid'' is an identifier unique
to a given astronomical object (e.g., 1-266039).  The second, the
``plate-ifu'' uniquely identifies a given observation by concatenating
the plate id with the IFU number (e.g., 7443-12701 identifies the
first 127-fiber IFU on plate 7443).  Since some galaxies are observed
more than once on different plates, a given mangaid can sometimes
correspond to more than one plate-ifu.

The mangaid consists of 2 parts separated by a hyphen. The first part
is the id of the parent catalog from which a target was selected. The
second part is the position within that catalog. For most galaxy
targets the catalog id is 1 which refers to the NSA. For a small
number of the early targets the catalog id is 12 and refers to an
earlier version of the NSA (v1b\_0\_2). All galaxies from this earlier
version of the NSA are also in the final version and so we release
photometry etc for those targets from the final version of the NSA
(v1\_0\_1), which is included in the data release. Other catalogs
referred to in the first part of the manga-id are for SDSS standard
stars.

The full data model for all MaNGA DRP data products can be found
online at \url{http://www.sdss.org/dr13/manga/manga-data/data-model/} and is also
given in Appendix B of L16.

\subsection{Retrieving MaNGA Data}

The raw data, reduced data, RSS, and 3-D data cubes for 1351 MaNGA
galaxies are provided in DR13. From these data products, maps of line
ratios, spectroscopic indices, and kinematics can be made using
standard software. Because the first step in using the MaNGA data for
science is to retrieve the spectra, we detail here and on the SDSS
website\footnotemark[4]\footnotetext[4]{\url{http:www.sdss.org/dr13/manga/manga-data/data-access/}} how to access the
MaNGA spectra.

\subsubsection{Reduced Data Products}

MaNGA DR13 reduced data products are stored on the Science Archive
Server at \url{http://data.sdss.org/sas/dr13/manga/spectro/redux/v1\_5\_4/}.  
The top level directory contains the summary drpall FITS table and
subdirectories for each plate.  Inside each plate directory there are
subdirectories for each MJD on which the plate was observed,
containing intermediate (exposure level) data products.  The `stack'
subdirectory within each plate directory contains the final RSS and
cube files for each MaNGA galaxy, sorted according to their plate-ifu
identifiers.  Note that the ifu identifier in the filenames indicates
the size of the IFU; everything in the 127 series (e.g., 12701) is a
127-fiber bundle, etc.  The 700 series ifus (e.g., 701) are the twelve
spectrophotometric 7-fiber minibundles that target stars on each
plate.

These are the ways of getting at the data in DR13:

\begin{itemize}
\item Direct html browsing of the SAS at the above link.  The file
  drpall-v1\_5\_4.fits can be downloaded through the web browser and
  queried using any program able to read FITS binary tables.  Once a
  set of galaxies of interest has been identified, individual data
  cubes, summary RSS files, intermediate data products, etc. can be
  downloaded by browsing through the web directory tree.
\item Large downloads of many DRP data products can be automated using
  rsync access to the SAS.  For instance, to download all MaNGA data
  cubes with a logarithmic wavelength format: \url{\\ rsync -aLrvz --include
  "*/" --include "manga*LOGCUBE.fits.gz" --exclude "*"
  rsync://data.sdss.org/dr13/manga/spectro/redux/v1\_5\_4/ ./}
\item The MaNGA drpall file can also be queried online using the SDSS
  CASJobs system at \url{http://skyserver.sdss.org/casjobs}.  While this can be useful for identifying
  MaNGA observations of interest, CASJobs does not contain links to
  the MaNGA data cubes and another method must be used to download the
  data themselves.
\item The SDSS SkyServer Explore tool at \url{http://skyserver.sdss.org/dr13/en/tools/explore/} will display basic
  information about MaNGA observations in DR13 that fall within a
  given cone search on the sky.  The relevant explore pages also
  provide direct links to the FITS data cubes on the SAS.

\end{itemize}

\subsubsection{Raw Data}

All MaNGA data taken in the first year of SDSS-IV observations are
part of Data Release 13, including data from special ancillary plates
and co-observing during APOGEE-2 time that are not part of the DRP
results. The raw data are stored on the SAS in the directory
\url{http://data.sdss.org/sas/dr13/manga/spectro/data/} in subdirectories
based on the MJD when the data were taken. The mangacore
directory\footnotemark[5]\footnotetext[5]{\url{http://svn.sdss.org/public/repo/manga/mangacore/tags/v1\_2\_3/}}
contains the information needed to figure out the RA and Dec positions
of fibers on plates, the dithered MJDs to be combined to make the
final spectrum in apocomplete directory, and information on the
calibration files. L16 and the
\url{http://www.sdss.org/dr13/manga/manga-data/metadata/} website contain
the relevant information about the file formats and the use of the
calibration files to get to the fully reduced spectra. Because these
files are prepared for internal use, they retain many old features
that should be ignored, such as the names assigned to targeting bits,
which still retain the names from SDSS-I.

\subsection{Notes on using MaNGA data}

There are several important caveats to keep in mind when working with
MaNGA data. In this discussion we treat only the MaNGA data
cubes. Summary RSS, intermediate, and even raw data products present
some statistical advantages to the data cubes, in particular reduced
covariance between adjacent data points and greater ease of forward
modeling, but are substantially harder to use.

First and foremost, each MaNGA data cube has a FITS header keyword
DRP3QUAL indicating the quality of the reduction. 1-2\% of the data
cubes are flagged as significantly problematic for various reasons,
ranging from poor focus to flux calibration problems. Table B13 in L16
lists the bits that can be set with this flag that describe why the
end product was deemed unsatisfactory. Galaxies for which the CRITICAL
quality bit (=30) is set in DRP3QUAL should be treated with extreme
caution. While there may be some spaxels in that data cube that are
acceptable, in general the CRITICAL bit indicates widespread problems
with the data reduction. Each data cube also has an associated MASK
cube describing the quality of individual spaxels in the data
cube. This MASK extension can be used to identify areas of no coverage
outside the fiber bundle footprint\footnotemark[6]\footnotetext[6]{The fiber footprint is a
  hexagon, but the standard FITS image data structure is based around
  rectangular arrays. There must therefore be blank areas around the
  live IFU footprint in order to inscribe the hexagon inside a
  bounding rectangle.}, low coverage near the edge of the dithered
footprint, problematic areas due to detector artifacts, regions known
to contain bright Milky Way foreground stars, etc. A simple summary
{\tt DONOTUSE} bit is of particular importance indicating elements
that should be masked out for scientific analyses.

Since the MaNGA data cubes adopt a 0.5$\arcsec$ sampling size (chosen
based on Fourier analysis in optimal observing conditions to not
truncate the high-k modes of the observational PSF), while individual
fibers have a 2$\arcsec$ diameter footprint, there is significant
covariance between adjacent MaNGA spaxels that must be taken into
account whenever combining spectra. A simplified method for doing
this is discussed in \S 9.3 of L16. The typical reconstructed point
spread function of the MaNGA data cubes has FWHM of 2.5$\arcsec$.

As discussed by L16, the instrumental line spread function (LSF) in
the wavelength direction reported by the various extensions within the
MaNGA DR13 data products is underestimated by about $10\pm2$\%.
This correction is comparable to the errors in the reported resolution of the BOSS spectrographs seen in single-fiber work \citep[e.g.,][]{2012AJ....143...90S,2013A&A...559A..85P}.
Although this makes little difference when determining the
astrophysical width of broad spectral lines, it is important to
account for when attempting to subtract the instrumental resolution
from barely-resolved lines. There are two effects that combine to produce this
overestimate. The first is that the impact of the wavelength
rectification on the effective spectral resolution was not accounted
for when combining spectra from blue and red cameras. The second is
that the Gaussian width of the LSF reported by the DRP is strictly the
width of a pre-pixellization gaussian, while most end-user analysis
routines adopt post-pixellization gaussians instead (i.e., the
different between integrating a gaussian profile over the pixel
boundaries vs evaluating a gaussian at the pixel midpoint). This will
be treated more completely in a future data release; in the meantime a
10\% correction to the instrumental LSF quoted by the MaNGA data
products appears to be a reasonable correction factor if using
post-pixellization analysis routines. However, because this correction
factor itself is uncertain, derived astrophysical line-widths
substantially below the instrumental resolution should be viewed to
have unreliable accuracy at this time. A full discussion of issues to 
consider is available at \url{http://www.sdss.org/dr13/manga/manga-caveats/}.

\subsection{Highlights of MaNGA Science with DR13 Data}
The MaNGA survey has produced a number of scientific results based on
early data, indicating the breadth of research possible with the MaNGA
data. Here we briefly summarize the results of papers completed within
the SDSS-IV collaboration using MaNGA data on spatially resolved gas
physics, stellar population properties, and stellar and gas
kinematics. These MaNGA papers serve as a guide to prospective users of
SDSS-IV data in these specific science areas. The larger context for this
science is provided to a small degree in the Introduction, and the papers here also provide
citations to the extensive literature on each topic, to which we refer
the interested reader for additional context. Citations to the science
in each area should refer to the original papers, whether MaNGA-based or
not, rather than this brief executive summary.

\subsubsection{Gas physics}

The spatially resolved emission-line measurements have been used to
understand the physical conditions of the ionized gas in
galaxies. \citet{2016Natur.533..504C} identified AGN winds as a
surprisingly common occurrence in normal, quiescent galaxies,
suggesting these winds as potentially critical in suppressing star
formation. These winds may help address the question of how star
formation remains suppressed in early-type galaxies.
\citet{2016Natur.533..504C} report bisymmetric emission features
co-aligned with strong ionized-gas velocity gradients in a galaxy from
which they infer the presence of centrally driven winds in typical
quiescent galaxies that host low-luminosity active nuclei.  These
galaxies account for as much as ten per cent of the quiescent
population with masses around $2\times 10^{10}\ M_{\odot}$. They
calculate that the energy input from the galaxies' low-level active
supermassive black hole is capable of driving the observed wind, which
contains sufficient mechanical energy to heat ambient, cooler gas
(also detected) and thereby suppress star formation.

The broader nature of ionized gas in early-types has also been the
subject of papers by \citet{2016MNRAS.461.3111B} and
\citet{2017MNRAS.466.2570B} following up on the analysis of a small
sample observed with the MaNGA prototype instrument in
\citet{2015MNRAS.449..867B}.  By using spatially resolved maps of
nebular diagnostics {\it and} stellar population ages, this work has
added substantial support to the notion that evolved stellar
populations provide the ionization source for a galaxy class that
arguably should be renamed from LINER (Low Ionization Nuclear Emission
Region) galaxies to LIER galaxies.
LIERs, it turns out, are ubiquitous in both quiescent galaxies and in
the central regions of galaxies where star formation takes place at
larger radii. The study of \citet{2016MNRAS.461.3111B} and \citet{2017MNRAS.466.2570B} 
have put the occurrence of the LIER phenomenon into a
physically relevant framework that can be directly tied to the
diversity of the galaxy population as a whole.  Specifically, they
identify two classes of galaxies as extended LIER (eLIER) and central
LIER (cLIER), respectively, and study their kinematics and stellar
population properties. cLIERs turn out to be mostly late type galaxies
located around the green valley, while eLIERs are morphologically
early types and are indistinguishable from passive galaxies devoid of
line emission in terms of their stellar populations, morphology and
central stellar velocity dispersion.

The widespread ionization state of LIER gas might originally manifest
as the Diffuse Intergalactic Gas (DIG) which is intermixed with
star-forming regions. Zhang et~al. (2016) studied galactic DIG emission
and demonstrate how DIG in star-forming galaxies impacts the
measurements of emission line ratios at the spatial resolution of
MaNGA, hence the interpretation of diagnostic diagrams and the
gas-phase metallicity measurements. They quantify for the MaNGA data
how the contamination by DIG moves HII regions towards composite of
LIER-like regions. DIG significantly biases measurements of gas
metallicity and metallicity gradients because at different surface
brightness, line ratios and line ratio gradients can differ
systematically.

A careful treatment of gas-phase metallicities inferred from spectral
maps of galaxies has suggested a key role for the dependence of
metallicity on local stellar mass surface density. \citet{2016MNRAS.463.2513B} 
present the stellar surface mass density vs gas
metallicity relation for more than 500,000 spatially-resolved
star-forming regions from a sample of 653 disk galaxies. These
galaxies span a larger range in mass than in previous samples where
the correlations were first discovered.  They confirm a tight relation
between these local properties, with higher metallicities as the
surface density increases. They find that even over this larger mass
range this local relationship can simultaneously reproduce two
well-known properties of disk galaxies: their global mass-metallicity
relationship and their radial metallicity gradients. Their results
support the idea that in the present-day universe local properties
play a key role in determining the gas-phase metallicity in typical
disk galaxies.

However, \citet{2016ApJ...832..182C} have found a galaxy in the
middle of a gas accretion event, providing a detailed look at what
appears to be a relatively rare occurrence in the nearby Universe of this
mode of galaxy growth. They present serendipitous observations of a
large, asymmetric H$\alpha$ complex that extends $\sim8\arcsec$
($\sim6.3$ kpc) beyond the effective radius of a dusty, starbursting
galaxy. This H$\alpha$ extension is approximately three times the
effective radius of the host galaxy and displays a tail-like
morphology. This is suggestive of an inflow, which is consistent with
its relatively lower gas-phase metallicities and its irregular gaseous
kinematics.

\subsubsection{Stellar populations}

Spatially resolved stellar population properties and stellar growth
histories have been analyzed, following the analysis of a small sample
observed with the MaNGA prototype instrument by
\citet{2015MNRAS.449..328W} and \citet{2015ApJ...804..125L}. MaNGA has
explored the role of environment in shaping the radial distribution of
stellar ages and metallicities, with particular attention given to the
potential measurement systematics. Using different spectral fitting
techniques and complementary environmental metrics, both \citet{2017MNRAS.466.4731G}
 and \cite{2017MNRAS.465.4572Z} conclude
that any environmental signal in the average shape of gradients is
weak at best, with no obvious trends emerging in the initial MaNGA
data.

\citet{2017MNRAS.465..688G} studied the internal gradients of the
stellar population age and metallicity within 1.5 $R_e$ obtained from
full spectral fitting and confirm several key results of previous
surveys. Age gradients tend to be shallow for both early-type {\it and} 
late-type galaxies. As well known from previous studies,
metallicity gradients are often complex (and {\it not} well-modeled by
linear or log-linear functions of radius), varying in detail from
galaxy to galaxy on small radial scales. On average, however, over
radial scales of order 1 $R_{\rm e}$, MaNGA data provide the strongest
statistical constraints to date that metallicity gradients are
negative in both early and late-type galaxies, and are significantly
steeper in disks. These results continue to indicate that the radial
dependence of chemical enrichment processes are far more pronounced in
disks than they are in spheroids, and indeed the relatively flat
gradients in early-type galaxies are inconsistent with monolithic
collapse. For both early and late-type galaxies, more massive galaxies
have steeper negative metallicity and age gradients.  Since early-type
galaxies tend to be more luminous, the overall steeper age and
metallicity gradients in late-type galaxies reflect the fact that the
trends in these gradients with galaxy mass are more pronounced for
late-type galaxies. \citet{2017MNRAS.465..688G} take advantage of
the unique MaNGA sample size and mass range to characterize this
correlation between metallicity and age gradients and galaxy mass,
which explains the scatter in gradients values seen in previous
studies.

\citet{2016MNRAS.463.2799I} meanwhile infer spatially-resolved
stellar mass assembly histories for the MaNGA galaxies, extending
previously known relations between galaxy type and assembly history to
a larger mass range. Their findings are consistent with
blue/star-forming/late-type galaxies assembling, on average, from
inside to out. Red/quiescent/early-type galaxies present a more
uniform spatial mass assembly, or at least one that has been
dynamically well mixed since star-formation ceased, consistent with
the flatter gradients seen, e.g., \citet{2017MNRAS.465..688G}.
In general, low-mass galaxies show evidence of more irregular global
and spatial assembly histories than massive galaxies.

In a developing effort to model stellar population gradients, \citet{2017MNRAS.465.2317J} 
demonstrate a new technique using MaNGA data
that seeks to decompose the underlying population into contributions
from different physical sub-components. They explore how the disk and
bulge components in galaxies reached their current states with a new
approach to fit the two-dimensional light profiles of galaxies as a
function of wavelength and to isolate the spectral properties of these
galaxies' disks and bulges. The MaNGA data have a spatial sampling of
0.5 $\arcsec$ per pixel, and successful decompositions were carried
out with galaxies observed with the 61- to 127-fiber IFUs with fields
of view of 22$\arcsec$ to 32$\arcsec$ in diameter respectively.

Rembold et al. (2017, submitted) have identified a ``control sample'' to the 
MaNGA luminous AGN host galaxies, matched on mass, distance, morphology, and
inclination. Their conclusions based on SDSS-III spectra of the central region
can be tested with evaluation of the stellar populations throughout the
galaxies with MaNGA data. 

\subsubsection{Gas and stellar kinematics}

Several studies are investigating the kinematics of both stars and gas
across the galaxy population. \citet{2017ApJ...838...77L} perform
dynamical modeling on a more extensive and diverse
sample of elliptical and spiral galaxies than had previously been
done. By comparing the stellar mass-to-light ratios estimated from
stellar population synthesis and from dynamics, they find a similar
systematic variation of the initial mass function (IMF) to previous
investigations. Early-type galaxies (elliptical and lenticular) with
lower stellar mass-to-light ratios or velocity dispersions within one
effective radius are consistent with a Chabrier-like IMF while
galaxies with higher stellar mass-to-light ratios or velocity
dispersions are consistent with a more bottom heavy IMF such as the
Salpeter IMF.

Two studies have taken advantage of the large MaNGA sample in DR13 to
quantify the frequency and attributes of galaxies with strong
disparities between gas and stellar kinematics. \citet{2016NatCo...713269C} 
find that an appreciable fraction of galaxies have counter-rotating gas and stars. 
Counter-rotation is found in about 2\% of all
blue galaxies. The central regions of blue counter-rotators show
younger stellar populations and more intense star formation than in
their outer parts. \citet{2016MNRAS.463..913J} have further studied
the properties of 66 galaxies with kinematically misaligned gas and
stars. They find that the star-forming misaligned galaxies have
positive gradients in D$_n$4000 and higher gas-phase metallicity,
while the green valley/quiescent ones have negative D$_n$4000
gradients and lower gas-phase metallicity on average. Despite this
distinction, there is evidence that all types of
kinematically-misaligned galaxies tend to be in more isolated
environments. They propose that misaligned star forming galaxies
originate from gas-rich progenitors accreting external kinematic
decoupled gas, while the misaligned green valley/quiescent galaxies
might be formed from accreting kinematic decoupled gas from dwarf
satellites.

Finally, \citet{2016MNRAS.462.3955P} examine the kinematics of a
sample of 39 quenched low-mass galaxies. The majority (87\%) of these
quenched low mass galaxies exhibit coherent rotation in their stellar
kinematics, and a number host distinct disks or spiral features. Just
five (13\%) are found to have rotation speeds $v_{\rm circ} < 15$ km
s$^{-1}$ at $1\ R_e$, and are dominated by pressure support at all
radii. Two of the quenched low mass galaxies (5\%) host kinematically
distinct cores, with the stellar population at their centers
counter-rotating with respect to the rest of the galaxy. The results
support a picture in which the majority of quenched low mass (dE)
galaxies have a disk origin.
\section{SEQUELS: eBOSS, TDSS, and SPIDERS data}

DR13 includes the data from 126 plates observed under the SEQUELS program. This program was started in SDSS-III as an ancillary program to take advantage of some of the dark time released when BOSS was completed early. The SEQUELS targets were quite different from BOSS targets because the program was designed to finalize the eBOSS target selection algorithms.  The primary targets were defined by two different SDSS-WISE selection algorithms to determine the eBOSS LRG program \citep{2016ApJS..224...34P} and several optical-IR and variability selections to determine the eBOSS quasar program \citep{2015ApJS..221...27M}. Likewise, objects selected from a combination of X-ray and optical imaging were used to determine the final SPIDERS cluster \citep{2016MNRAS.463.4490C} and AGN \citep{2017arXiv170401796D} programs while variability in PanSTARRS imaging was used to determine the final TDSS program \citep{morganson2015,ruan2016}. 

66 SEQUELS plates were completed in the final year of SDSS-III.  The remaining 60 plates required to fill out the 400 square degree footprint were completed in the first year of SDSS-IV.  As mentioned above, these data served a crucial role for verification of the eBOSS, TDSS and SPIDERS target samples.  SEQUELS and eBOSS LRG spectra were used to optimize the performance of a new redshift classification scheme that now reliably classifies more than 90\% of eBOSS LRG spectra \citep{2016AJ....152..205H}, thus meeting the ambitious goal set forth at the beginning of the program \citep{2016AJ....151...44D}.  The sample also seeds the eBOSS footprint to be used for clustering studies.  The first clustering measurements from SEQUELS and eBOSS LRGs were just released \citep{2016arXiv160705383Z} and first results from quasar clustering are expected in the near future. All SEQUELS targets are tracked by the {\tt EBOSS\_TARGET0} bitmask.  The appendix of the DR12 paper \citep{2015ApJS..219...12A} provides the motivation and description of each target selection algorithm captured by that bitmask.

\subsection{eBOSS in SEQUELS}

117 plates from SEQUELS used a slightly broader selection for LRG, clustering quasars ($0.9<$z$<2.2$), and Lyman-$\alpha$
forest quasars to ensure that the final eBOSS selection would be included in each of these classes. Because the eBOSS
sample is included in this region, the SEQUELS spectroscopy obtained in SDSS-III and SDSS-IV will be used directly in
any LRG or quasar clustering studies.  Nine plates from SEQUELS, all released in DR12, included targets derived from an early test of the ELG
selection algorithm \citep{2015ApJS..219...12A}. These tests of ELG selection algorithms were part of a larger series of tests performed during
SDSS-III and SDSS-IV (\citealt{comparat,raichoor,delubac2017}. The spectra from these tests
were also used in one of the first science results from eBOSS, a study of galactic-scale outflows traced by UV emission
\citep{2015ApJ...815...48Z}. The selection algorithm used in these fields is quite different from what will be used in eBOSS and
these targets will not contribute to the final clustering sample. For this reason, we summarize the findings of the LRG
and quasar spectra below but reserve discussion of the ELG spectra for future work.

DR13 does contain value-added catalogs relevant to ELG sample. \citet{raichoor} describe
the Fischer discriminant used to select ELG targets using photometry from SDSS, WISE, and SCUSS \citep{2016AJ....151...37Z} for the main ELG sample. 
\citet{delubac2017} produce the catalogs used for ELG targeting with SDSS+WISE or SDSS+WISE+SCUSS data. Finally, \citet{2015ApJ...815...48Z} generated
composite, continuum-normalized, spectra of emission-line galaxies in the rest-UV from ELGs observed by BOSS in various programs.
Table~2 lists the location of these files on the SAS.

\subsubsection{Luminous Red Galaxies from WISE colors}

 The increase in redshift complicates selection both by shifting the 4000 \AA\ break into the $i$-band filter and by requiring fainter targets than those observed in BOSS.  
 WISE 3.4 $\mu$m photometry (W1 band) was introduced to enable selection of this higher redshift sample.  
As part of the SEQUELS program, two overlapping selections for LRGs at higher redshifts ($0.6
< z < 1.0$, vs $0.4 < z < 0.7$ for CMASS) were employed, allowing tests of potential
strategies for eBOSS. Color cuts that combine optical and infrared photometry were
employed, enabling the targeting of LRGs at these redshifts while maintaining a high purity.
This selection technique takes advantage of the strong peak at a rest frame wavelength of 1.6
microns that is present in the spectrum of most galaxies.  This peak enters W1 as the redshift gets closer to 1, 
yielding large differences between the optical/IR colors of $z>$0.6 galaxies and stars.

SEQUELS selected a total of $\sim 70,000$ LRGs over an area of $\sim 700$ deg$^2$  with $120.0
< $RA$ <  210.0$ and $45.0 <$ DEC $<  60.0$. These LRGs were selected by algorithms utilizing two
different optical-IR color spaces, utilizing either SDSS $r$, $i$, $z$, and W1, or only $i$, $z$ and W1; 
the selection efficiency and redshift success for each algorithm could then be compared. The parameters of
the selection algorithms were tuned such that each yielded a target density of $\sim 60$
deg$^{-2}$; the two selections overlap significantly, yielding a net density of 72 targets
deg$^{-2}$. Figure 10 of  \citet{2016ApJS..224...34P} presents the resulting redshift distributions
from each selection. The $r$/$i$/$z$/WISE selection has been chosen for eBOSS due to
greater efficiency at selecting high-redshift LRGs. More details on the SEQUELS LRG sample
selection can be found in \citet{2016ApJS..224...34P}.

\subsubsection{Quasars targeted with Optical+WISE photometry and photometric variability}
SEQUELS observations helped define a uniform quasar sample for eBOSS clustering studies based on $ugriz$ and WISE photometry.  The ``Extreme Deconvolution" \citep[XDQSO;][]{bovy2011a,bovy2011b} selection is used to identify likely quasars at redshifts $z>0.9$ according to the relative density of stars and quasars as a function of color, magnitude, and photometric uncertainty. The selection alone results in a highly complete sample of quasars to be used for clustering studies, but with contamination from stars that is too large to fit into the eBOSS fiber budget.   The XDQSO selection is supplemented by morphology cuts to remove low redshift AGN and optical-IR colors to remove stellar objects with blackbody spectra. Variability data from the Palomar Transient Factory \citep[PTF;][]{2009PASP..121.1395L}  are used to supplement the selection of Lyman-$\alpha$ forest quasars, producing tracers with a density of 3.2 deg$^{-2}$ where sufficient PTF imaging data are available. In addition to cosmological measurements, the quasar sample can be used to study quasar astrophysics and galaxy evolution through studies of the quasar luminosity function, composite spectra, and multi-wavelength spectral energy distributions spanning the radio to the X-ray. \citet{2015ApJS..221...27M} found that $\sim$ 96\% of all quasar targets with $r < 22$ will be confidently classified. Section 5 of  \citet{2015ApJS..221...27M} provides information on the properties of  quasars observed with SEQUELS, including the numbers observed in the redshift ranges  $0.9< z <2.1$ and $z > 2.1$.

DR13 includes value-added catalogs with the variability measurements \citep[see][]{2011A&A...530A.122P} using either PTF data \citep{2015ApJS..221...27M} or Stripe 82 data \citep{qsovar} that
are used for selecting quasars in eBOSS based on variability. The locations of the FITS tables are given in Table~2.

\subsubsection{Redmonster and Improved Redshifts for LRGs}
The DR13 redshifts for all SEQUELS targets are determined in the usual fashion, with best-fitting combinations of PCA eigenspectra. \citet{2016AJ....152..205H} describe a new pipeline, {\tt redmonster} that uses a suite of discrete theoretical galaxy spectra as a basis to determine the most likely redshift through a minimum $\chi^2$, rather than linear combinations of templates used in  \citet{2012AJ....144..144B}. This pipeline achieves a 90.5\% automated redshift and spectral classification success rate for the LRG target class, a significant improvement over the performance of the previous pipeline. A value-added catalog using the new {\tt redmonster} algorithms for the LRG sample is included in DR13. All spectra identified by {\tt EBOSS\_TARGET0} bit 1 or 2  were classified by redmonster.  The file is named {\tt redmonsterAll-v5\_9\_0.fits} and is found on the SAS as described in Table 2. 

\subsection{SPIDERS in SEQUELS}

The main goal of the SPIDERS survey is to characterize a subset of X-ray sources identified by eROSITA using optical spectra from the BOSS spectrograph. The extended sources will mostly be galaxy clusters, which can be used for cosmology. The point sources will mainly be AGN, which can be used to study the evolution of black holes across cosmic time. For the first phase of SDSS-IV, when eROSITA data are not yet available, SPIDERS will be targeting based on ROSAT and XMM data. The target catalogs for galaxy clusters and AGN for SPIDERS from these two satellites have been included in DR13 as value-added catalogs. The SPIDERS AGN target catalogs \citep{2017arXiv170401796D} contain 9,028 candidate targets from RASS and 819 from XMMSL \citep[{\bf XMM}-Newton {\bf Sl}ew survey catalog;][]{2008A&A...480..611S}. They enclose information on the X-ray sources, including flux measurements, and a quantitative measure of the reliability of the association to optical and AllWISE data. The SPIDERS Galaxy Cluster target list \citep{2016MNRAS.463.4490C} contains 94,883 and 3,839 objects for RASS and XMM respectively. 

In SEQUELS, SPIDERS used similar targeting catalogs, also available as value-added products, to test targeting strategies and provide initial results. The selection criteria are somewhat different than the final SPIDERS algorithms. Full details are available in \citet{2016MNRAS.463.4490C} and \citet{2017arXiv170401796D}. We summarize the SEQUELS SPIDERS data available in DR13 below.

\subsubsection{Optical spectra of Galaxies in X-ray Identified Clusters}

\begin{figure*}[htb]
\begin{center}
\includegraphics[width=5in]{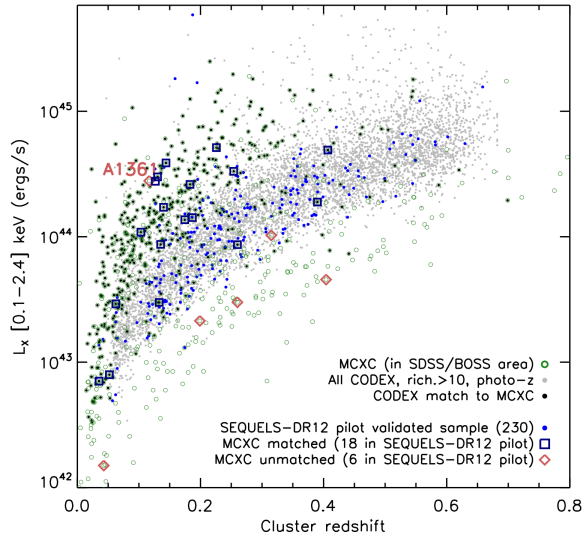}
\caption{{ 
Distribution in the X-ray luminosity-redshift plane of galaxy clusters, adapted from \citet{2016MNRAS.463.4490C}. The gray dots mark the location of all candidate CODEX clusters using their photometric redshift. They form the main pool of cluster targets in SPIDERS. The CODEX candidate clusters that have already been spectrocopically confirmed and included in the MCXC meta-catalog \citep{piffaretti} are indicated by open symbols. The confirmed clusters lie mainly at higher X-ray luminosities. The 230 confirmed clusters in the SPIDERS DR13 value-added catalog  (blue dots) extend the redshift range of known clusters in the SDSS area to $z\sim0.6$ in a systematic way.  Among these 230, only 18 match clusters from the MCXC meta-catalog. The new clusters in general lie at lower X-ray luminosities and therefore probe lower X-ray masses than previous ROSAT-based analyses. The 6 clusters indicated by red diamonds are MCXC clusters in the SEQUELS-DR12 footprint not present in SPIDERS (note that ABELL 1361 is within a masked area of the CODEX survey, so it is not matched). 
}}
\label{xrayclusters}
\end{center}
\end{figure*}

The cluster pilot study \citep{2016MNRAS.463.4490C} takes advantage of the CODEX ({\bf Co}nstrain {\bf D}ark {\bf E}nergy with {\bf X}-ray clusters; Finoguenov et al., in preparation) candidate cluster list, which is based on currently available RASS data. As part of DR13, we provide the catalog of X-ray detected galaxy clusters spectroscopically confirmed using SEQUELS-DR12 SPIDERS spectroscopic data. Galaxy clusters were identified through the emission of their hot baryonic component as extended X-ray sources in RASS. The optical counterparts were found by optimally searching photometric data for the existence of a red-sequence formed by their member galaxies. Spectroscopic redshifts obtained by SPIDERS provide definitive confirmation of the clustered nature of these objects and their redshift (up to $z\sim0.6$). We assigned cluster membership combining an algorithm and visual validation of individual objects. The gas properties derived from X-ray observations (luminosity, temperature, R$_{200}$\footnotemark[7]\footnotetext[7]{R$_{200}$ is defined as usual as the radius where the mean overdensity is equal to 200$\times$ the critical density.}) are derived using precise cluster redshifts ($\Delta_{z}  \sim 0.001$). We compute galaxy cluster velocity dispersions using several methods adapted to the low number of spectroscopic members per system (of the order 15-40) and we show that their values correlate with cluster X-ray luminosities, within expectations. Figure~\ref{xrayclusters} shows the distribution of clusters with SEQUELS redshifts and membership in the redshift-X-ray luminosity plane. 

\subsubsection{Optical spectra of X-ray-identified AGN}
The addition of X-ray-identified AGN to the suite of AGN with well-sampled redshifts helps complete the inventory of AGN and trace black hole growth throughout cosmic history. SDSS has been observing optically-identified AGN since its inception under the main large-scale structure surveys, special targeting programs, and as ``mistakes'' in other targeting classes. In addition, there were several BOSS ancillary programs focused on X-ray follow-up, including a highly complete program on the XMM-XXL north field \citep{2016MNRAS.457..110M}. The SPIDERS SEQUELS program has added spectra of 274 ROSAT AGN to the SDSS sample, identified on the basis of their SDSS colors only. The DR12 paper (Alam et al. 2015) and \citet{2017arXiv170401796D} provide details on the targeting of these AGN.

\subsection{TDSS in SEQUELS}

\begin{figure*}[htb]
\begin{center}
\includegraphics[width=5in]{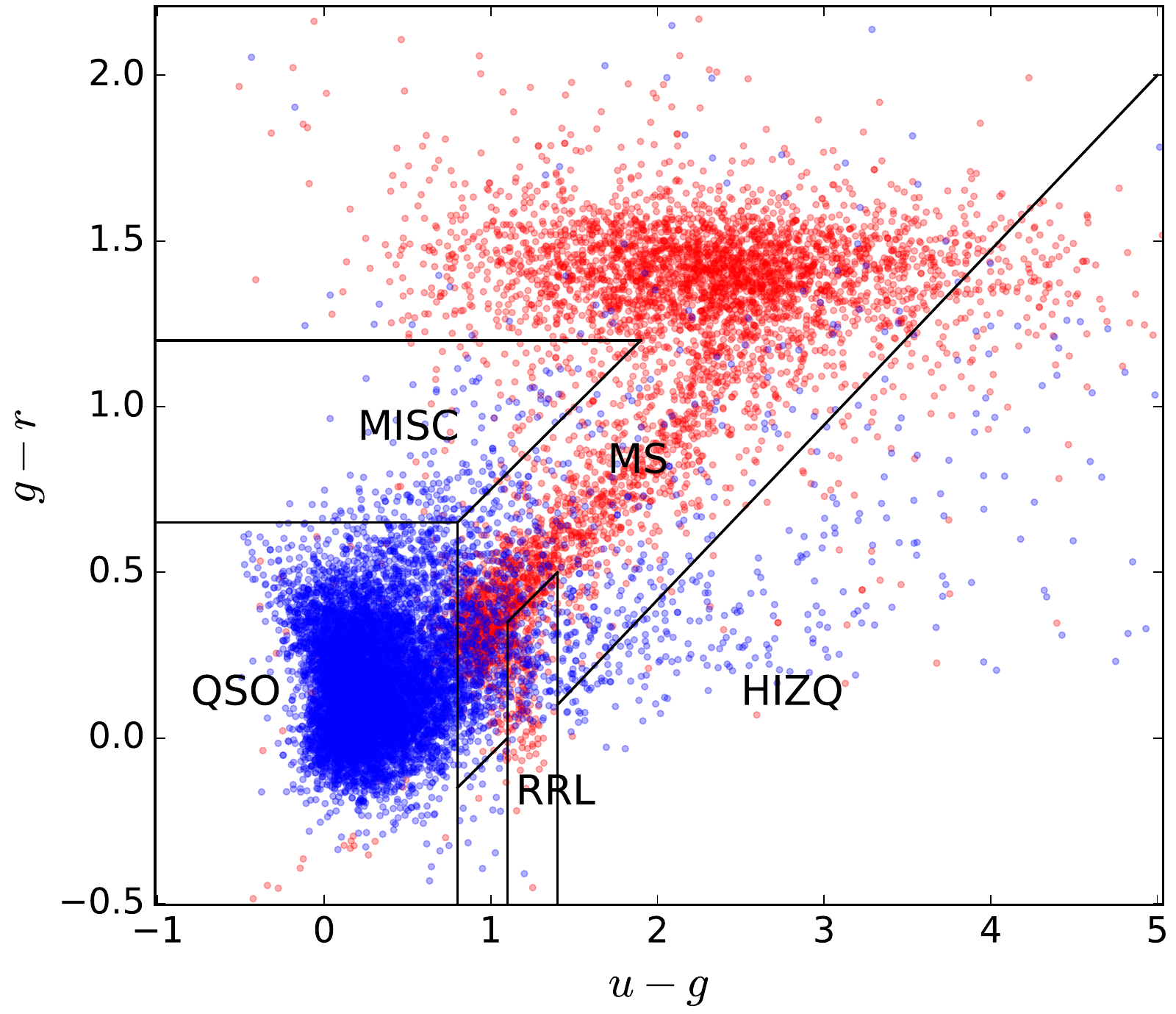}
\caption{{ A representation of the TDSS spectroscopic characterization of imaging variables selected (from PanSTARRS 1 and SDSS imaging) without any primary bias based on color or specific lightcurve character. Nearly 16,000 TDSS photometric variables in the region of the sky covered by the 66 initial SEQUELS plates are plotted here in a traditional color-color diagram. Their classifications are based on spectra taken in SDSS-I-III, including SEQUELS data. Regions in color space traditionally considered as occupied by quasars, main-sequence stars, RR Lyrae, high-redshift quasars, and other miscellaneous subclasses are overlayed  (see \citealt{sesar2007}, but here using specific boundaries detailed in \citealt{morganson2015}). The blue symbols depict TDSS photometric variables whose SDSS spectra verify them as quasars, while the red symbols depict TDSS variables whose SDSS spectra verify them as stars. A few hundred objects that were identified as galaxies or which could not be identified are not included in this plot. The wide distribution, as well as the overlap (in some regimes relatively densely), of both stars and  quasars symbols verify that TDSS variability selection finds not  only traditional quasars, but also those within color regimes commonly  attributed to stars (and with analogous results for the spectroscopically confirmed variable stars).
}}
\label{tdss}
\end{center}
\end{figure*}

Nearly 18,000 targets selected or co-selected by TDSS have been observed among the 126 SEQUELS spectroscopic plates. The targeting strategy for TDSS in SEQUELS was very similar to that ultimately chosen for the bulk of SDSS-IV \citep{morganson2015}. \citet{ruan2016} present TDSS spectroscopic results from the 66 initial DR12 SEQUELS plates, along with a description of the small differences in targeting within SEQUELS. Figure~\ref{tdss} depicts results for the initial TDSS  SEQUELS sub-sample. The classification of spectra was initially done using the BOSS pipeline \citep{2012AJ....144..144B}, but the spectra were also  visually inspected. Overall, the pipeline performance is outstanding, with the highest-level spectral classification (e.g., star vs. galaxy  vs. quasar) in agreement with our visual inspection for about 97\% of the TDSS spectra and with only 2\% of the pipeline redshifts for quasars requiring significant adjustment.

About 90\% of the TDSS spectroscopic fibers are aimed at initial  classification spectra for variables chosen without primary bias based on color or specific lightcurve character. Their variability is  determined from within PS1 multi-epoch imaging \citep{2010SPIE.7733E..0EK}, or via longer-term photometric variability between SDSS and PS1 imaging surveys \citep{morganson2015}. The initial SEQUELS results reveal comparable numbers of stars and quasars among these photometric variables. A summary for TDSS quasars is that the PS1/SDSS variability criteria mitigate some of the (well-known) redshift biases of color-selection yielding a smooth and broad quasar redshift distribution, and that TDSS selects relatively redder quasars (e.g., than the eBOSS core quasar sample) as well as a higher fraction of some peculiar AGN (such as BALQSOs, and BL Lacs); and among variable stars, TDSS selects significant numbers of active late-type stars, stellar pulsators such as RR Lyrae, and eclipsing and composite binaries, along with smaller subsets of white dwarfs, cataclysmic variables, and carbon stars (see \citealt{ruan2016}).

The other $\sim$10\% of the TDSS targets are objects already having one or more earlier epochs of SDSS I-III spectroscopy, and for which TDSS is taking a second (or sometimes 3rd or 4th etc.) epoch to reveal anticipated spectroscopic variability. These ``Few Epoch Spectroscopy" (or FES) targets include various subsets of quasars and stars \citep{morganson2015}. Recent example science papers representative of this FES category of TDSS, include work on acceleration of broad  absorption lines in BALQSOs (e.g., Grier et al. 2016), and recent  discoveries adding to the rare class of ``changing look quasars" \citep[e.g.,][]{runnoe2016}.

\subsection{Retrieving SEQUELS data}

All SDSS data releases are cumulative and therefore the SEQUELS data, whether taken in SDSS-III or SDSS-IV, have been reduced with the latest pipelines and included with all previous SDSS optical spectra data in this data release. SEQUELS targets can be identified because the {\tt EBOSS\_TARGET0} bit will be set. The summary spAll-v5\_9\_0.fits datafile, which includes classification information from the pipeline, is at \url{https://data.sdss.org/sas/dr13/eboss/spectro/redux/v5\_9\_0/} on the SAS or in the specObjAll table on the CAS.

\section{APOGEE-2: Improved Spectral Extraction and Spectroscopic Parameters for APOGEE-1 data}

The data released in DR13 are based on the same raw data as in DR12, and the pipelines for reduction and analysis remain similar to those used in DR13. First, the data reduction pipeline \citep{nidever2015} reduced the 3-D raw data cubes into well-sampled, combined, sky-subtracted, telluric-corrected and wavelength-calibrated 1-D spectra.  The stellar parameters and abundances were derived using ASPCAP \citep[{\bf A}POGEE {\bf S}tellar {\bf P}arameters and {\bf C}hemical {\bf A}bundances {\bf P}ipeline; ][]{ASPCAP} by finding the $\chi^2$ minimum when comparing the normalized observed spectra against a grid of synthetic spectra. However, the processing and analysis have been improved in several ways:
\begin{itemize}
\item The linelist used for determining stellar parameters and abundances has been revised.
\item Abundances are derived for several more species  (\ion{C}{1}, P, \ion{Ti}{2}, Co, Cu, Ge, Rb) than in DR12.
\item Results are now available for stars with T$_{\rm eff}<$ 3500 using a newly employed MARCS model atmosphere grid. 
\item Separate synthetic spectral grids are used for dwarfs and giants; results for dwarfs include rotation and different isotope ratios are used for dwarfs and giants. 
\item An initial, approximate, attempt has been made to account for variable line-spread functions (LSFs) as a function of fiber number.
\item The correction for telluric absorption has been improved, primarily from a better LSF characterization, leading to better recovery of the stellar spectra.
\item The relation adopted from microturbulence has been refined, and a relation for macroturbulence has been incorporated.
\end{itemize}

More details are provided in Holtzman et al (2017, in preparation;
H16). Overall, these changes have improved and enhanced the
APOGEE stellar parameters and abundances, but users need to continue
to be aware of potential issues and data flagging, as discussed on
the SDSS web site and H16. In particular, parameters and abundances
for low metallicity stars and for cool stars are more uncertain.

Subsequent to the freezing of the DR13 release, several issues were
discovered with the ``calibrated’’ effective temperatures and
surface gravities that are released with DR13. Based on good agreement
of the spectroscopic effective temperature with photometric effective
temperatures for the bulk of the sample that is near solar-metallicity,
no calibration was applied to the DR13 spectroscopic effective
temperatures. It now appears, however, that these are systematically
offset from photometric temperatures for stars of lower metallicity by
as much as 200-300 K for stars at [Fe/H]\footnotemark[8]\footnotetext[8]{[X/H]=(log$_{10}$(N$_{\rm X}$/N$_{\rm H}))_*-$(log$_{10}$(N$_{\rm X}$/N$_{\rm H}))_{\odot}$. [X/Fe]$=$[X/H]$_*-$ [Fe/H]$_*$.} 
$\sim-2$.  For surface gravities, a
calibration was derived based on asteroseismic surface gravities. While
this calibration is good for most of the sample, it now appears that
it does not yield accurate surface gravities for metal-poor stars; the
calibrated surface gravities for stars with [M/H]\footnotemark[9]\footnotetext[9]{[M/H] is closely related to [Fe/H]. See \citet{2013AJ....146..133M} for more details.}$<-$1.5 are systematically
too low, by as much as $\sim0.5$ dex. H16 and the SDSS web site discuss both of
these issues, and suggest post-release calibrations.

\subsection{Improvements in Linelist and Data Analysis}

Several improvements were made with regard to the APOGEE LSF.
The characterization of the LSF was improved in one of the detectors,
and an initial attempt was made at accommodating the LSF variability
in the parameter and abundance pipeline by constructing spectral
libraries for 4 different LSFs and using the most appropriate one
for the analysis of each star. 

The linelist adopted for DR13,
linelist.20150714, is an updated version of the one used for
DR12 results \citep{2015ApJS..221...24S} and is available at
\url{http://data.sdss.org/sas/dr13/apogee/spectro/redux/speclib/linelists/}.
\citet{2015ApJS..221...24S} noted a number of concerns with the DR12
linelist, which have been corrected in the new version (H16). As in
DR12, the molecular line list is a compilation of literature sources
including transitions of CO, OH, CN, C2, H2, and SiH. The CN line list
has been updated from the DR12 version using a compilation from C. Sneden
(private communication). All molecular data are adopted without change,
with the exception of a few obvious typographical corrections. The atomic
line list was compiled using a number of literature sources for the lower
excitation state of the transition and the transition probability, which is usually characterized
as the ``gf'' or ``log gf'' value.  Literature sources include 
theoretical, astrophysical, and laboratory gf values.  A few new lines were added from NIST5\footnotemark[10]\footnotetext[10]{\url{http://www.nist.gov/pml/data/asd.cfm}}
and other literature publication since the DR12 line list was created,
including hyperfine splitting components for Al and Co.  We still rely
heavily on ``astrophysical'' gf values for atomic lines, where
the transition probabilities of individual lines are determined using
the spectrum of a star with known parameters and abundances, to construct
linelists in the H-band \citep{melendez99,ryde2015}. For our linelist, we
use the center-of-disk spectrum of the Sun \citep{1991aass.book.....L}
and the full disk spectrum of the nearby, well-studied, red giant
Arcturus \citep{1995PASP..107.1042H}. To calculate the astrophysical
gf values, we used Turbospectrum \citep{ap1998,plez2012} to generate
synthetic spectra of the center-of-disk for the Sun and the integrated
disk for Arcturus with varying oscillator strengths and damping values
to fit the solar and Arcturus spectra. In DR12, a different stellar
atmosphere code was used for the gf determination and the synthetic
grid creation and synthetic integrated disk spectra were used for the
both the Sun and Arcturus. Although the change in gf values from these
changes is small \citep{2015ApJS..221...24S}, the DR13 linelist is more
self-consistently generated than previous versions.  In DR12, we derived
final astrophysical gf values by weighting the solar astrophysical gfs
at twice those of Arcturus. The astrophysical gf solutions in DR13 are
weighted according to line depth in Arcturus and in the Sun, to give
more weight to where the relatively weak lines in the H-band produce
the best signal, which usually gives more weight to the Arcturus solution.
For lines with laboratory oscillator strengths, the astrophysical log(gf) values were not
allowed to vary beyond twice the error quoted by the source.

\subsection{Additional Elements}

\begin{figure*}[htb]
\begin{center}
\includegraphics[width=5in]{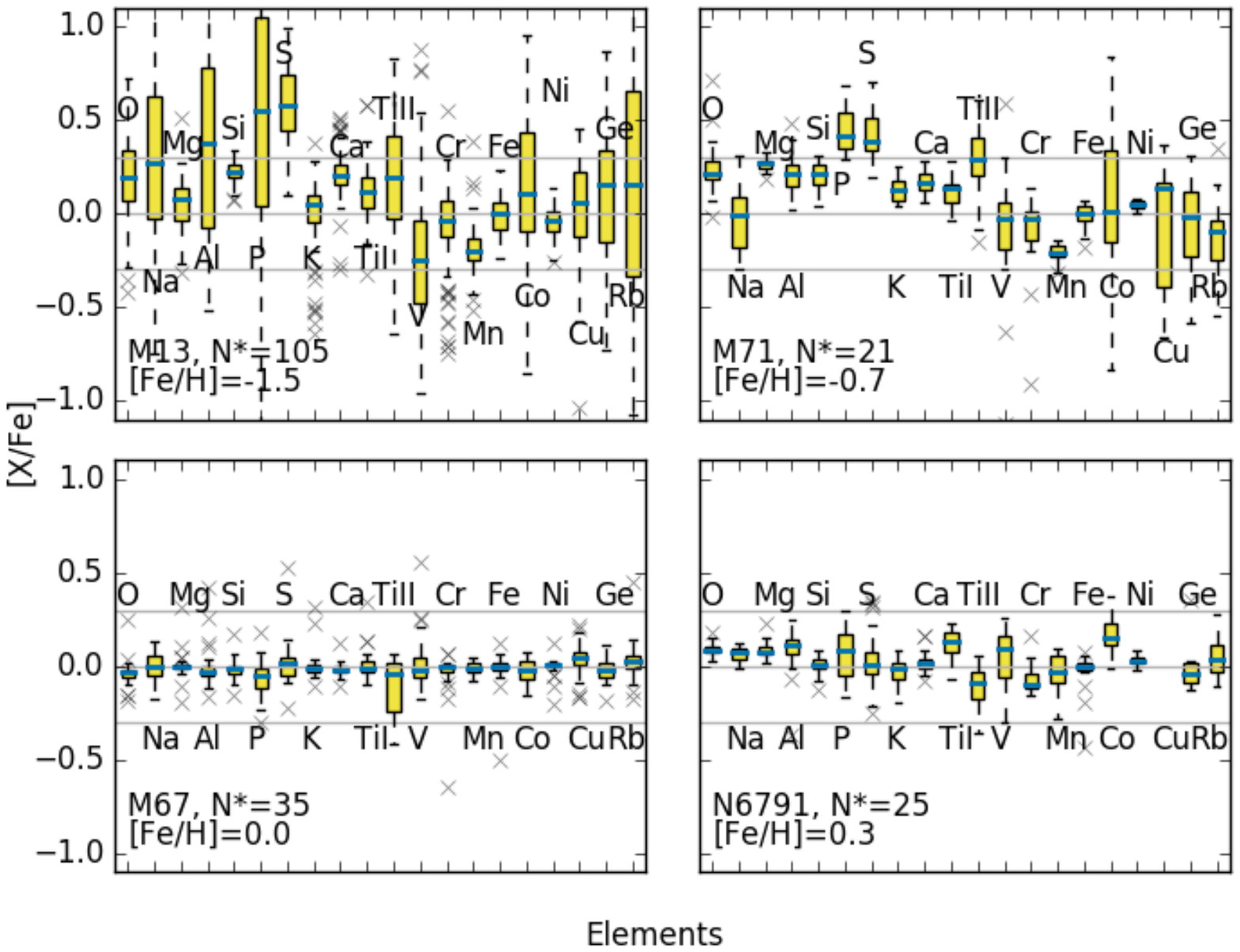}
\caption{A box plot of the stars with 4100K $<$ T$_{\rm eff} <$ 5000K in 4 clusters spanning a range in metallicities (M13$=-1.58$, M71$=-0.82$, M67$=0.06$, NGC~6791$=0.37$ \citep{holtz2015}. C and N are not included because their values in a cluster change depending of the evolutionary state of the star as the result of first dredge-up \citep[e.g.,][]{masseron2015}. A boxed horizontal line indicates the interquartile range 
and median found for a particular element. The vertical tails extending from the boxes indicate the total range of abundances determined for each element, excluding outliers. Outliers are shown by $\times$ symbols.}

\label{clusterabund}
\end{center}
\end{figure*}

In DR13, APOGEE provides abundances for elements P, Cr, Co, Cu, Ge, and Rb for the first time. The abundances of C, N, O, Na, Mg, Al, Si, S, Ca, Sc, Ti, V, Mn, Fe, and Ni were re-derived. We added two new species: \ion{C}{1} and \ion{Ti}{2}, which provide valuable checks on the abundances derived from CH and \ion{Ti}{1} lines (H16). The technique for calculating abundances of individual elements is described in \citet{holtz2015}. As for DR12 and as discussed in \citet{holtz2015}, we correct abundance ratios [X/Fe], except for [C/Fe] and [N/Fe], so there is no trend with temperature among the members of a single star cluster. In Figure~\ref{clusterabund}, we show box plots for the abundances for member stars in four clusters with a range of metallicity. Both the mean values and the rms illustrate key points about the APOGEE DR13 abundances.

\begin{itemize}
\item The elements cover a wide range in nucleosynthesis sites and atomic number. APOGEE is measuring the abundances of $\alpha$-elements, odd-Z elements, iron-peak elements, and neutron-capture elements, as well as the mass, mixing and AGB-contribution diagnostics C and N.
\item The high [X/Fe] for the $\alpha$ elements in the metal-poor ([Fe/H]$<-$0.7) globular clusters is present as expected \citep{1962ApJS....6..407W}. The increased scatter in O, Mg, and Al for these same clusters is in part due to the well-known light element anomalies in globular clusters \citep[e.g.,][]{kraft1994,meszaros_gcc}. The lower [Mn/Fe] in metal-poor stars is consistent with previous work \citep[e.g.,][]{1962ApJS....6..407W,1989A&A...208..171G} using 1-dimensional models that assume local thermodynamic equilibrium. 
\item The increased scatter at lower metallicities is not entirely
the result of actual inhomogeneities in cluster members, as there
are increasing uncertainties associated with the weaker lines in more
metal-poor stars. The error estimates reported in DR13 for these stars track the 
measured rms reasonably well, especially for errors in [X/Fe] $>$ 0.1 dex, and therefore provide a useful guide for abundance ratios that are not well-determined (H16).
On top of this, the ASPCAP methodology breaks down
at some level for stars in which abundances within a given ``abundance
group” (e.g., the alpha-elements) depart from solar abundance ratios, as
is the case for second-generation stars in metal-poor globular clusters \citep{meszaros_gcc}.

\item ASPCAP reports the parameters of the best-fitting model in $\chi^2$ space. It does not currently report non-detections and upper limits. Therefore, most reported abundances for elements such as P that show $\sim$1 dex scatter in Figure~\ref{clusterabund} are based on fits to noise in the spectrum and do not reflect actual abundances. Abundances for such elements should not be used unless they are confirmed by visual inspection. They are reported because for a certain range in metallicity and temperature, the lines can be detected and interesting chemistry exposed \citep[e.g.,][]{hawkins}.
\item The disagreement between the [Ti/Fe] value based on \ion{Ti}{1} lines and \ion{Ti}{2} lines is yet to be resolved. We are currently investigating the choice of lines (as noted by \citealt{hawkins}) and the effect of 3-D and NLTE corrections.
\item The mean and scatter shown in Figure~\ref{clusterabund} were calculated using stars with  $4100$K $<$ T$_{\rm eff} < 5000$K. The scatter becomes noticeably worse if the warmest or coolest stars in the clusters are included. Warmer stars have weaker lines in general in the H-band and the coolest stars are affected by the issues with the cool grid (see below).
\end{itemize}

The elements included in DR13 are by no means the only elements/species with lines present in the H-band amenable to abundance analysis. \citet{hawkins}, for example, independently derives Yb, along with many other elements, for the APOKASC sample \citep{apokasc}. We expect to include additional elements in upcoming data releases.

\subsection{Synthetic Spectral Grids at T$_{\rm eff} < 3500$K and with Rotational Broadening }

In DR10-12, the synthetic spectral grid used by ASPCAP \citep{meszaros2012,zamora2015} was restricted to
temperatures hotter than 3500K because Kurucz model atmospheres are not
available at cooler temperatures. However, many important APOGEE targets
have cooler temperatures, including the luminous metal-rich M-giants
in the bulge, cool asymptotic giant branch stars, and M dwarf planet
hosts. Therefore, we used custom MARCS \citep{marcs} atmospheres provided
by B. Edvardsson (private communication) to construct new synthetic
spectra. For the stellar atmospheres for the giants, the atmospheres
are spherical, otherwise they are plane-parallel. The Kurucz model grid
and the new MARCS model grid overlap between 3500-4000K. In this region,
there are some systematic differences between the results from the two
grids; to provide homogeneous results, DR13 adopts the parameters and
abundances from the analysis with the Kurucz grid.

Figure~10 shows the current parameter space in
T$_{\rm eff}$, log g and [M/H] covered by Data Release 13 stars. It is
immediately apparent that the parameters derived from the MARCS grid
do not match smoothly to the parameters from the Kurucz grid. Possible
explanations include the switch from plane-parallel Kurucz to spherical
MARCS for the giants and/or the large number of model atmospheres
in the MARCS grid that failed to converge. Because ASPCAP requires
a square grid of synthetic spectra, these ``grid holes'' were filled
by adjacent model atmospheres. The linelist does not have FeH lines,
which can be important features in the atmospheres of cool M dwarfs. We
are investigating the size of the error caused by using incorrect model
atmospheres in the grid, the possibilities of using alternative methods
of interpolating and identifying the best-fit model, and the addition of
FeH lines. H16 provides more details on the construction of the cool grid
and the resulting stellar parameters. Nonetheless, Figure~10
illustrates the improved parameter space over DR12, which should aid
in classifying M stars correctly as dwarfs or giants and separate the
early M from the late M stars.

\begin{figure*}[htb]
\label{hrdiagram}
\begin{center}
\includegraphics[width=5in]{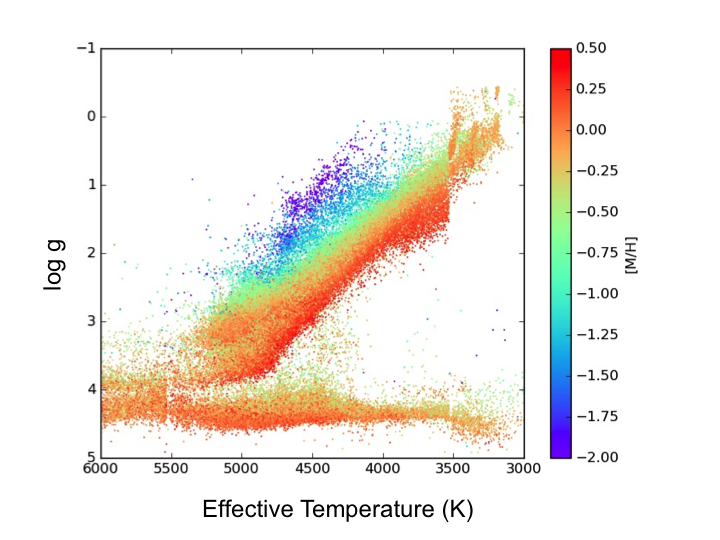}
\caption{The Hertzsprung-Russell diagram for all DR13 APOGEE-1 stars. The calibrated gravities and temperatures are used. The cool stars now clearly fall into dwarf and giant categories, but the stellar parameters do not merge smoothly onto the warmer Kurucz-based grid.}
\end{center}
\end{figure*}

The APOGEE-1 sample is dominated by giants used to probe the chemical cartography of the Galaxy \citep{Majewski_2017}. Fewer than 2\% of red giants rotate at speeds detectable at the APOGEE spectrograph resolution \citep{1996A&A...314..499D, 2011ApJ...732...39C}. Therefore the first versions of ASPCAP did not include rotational broadening as a dimension in the synthetic spectral grid, which substantially reduced the computing time. However, dwarf stars, of which there are $>$20,000 in APOGEE-1, are frequently rapidly rotating, especially if they are young. For DR13, we added a dimension to the synthetic spectral grid where spectra were broadened by a Gaussian kernel. To keep the computing time reasonable, and in acknowledgment of the small effect that C and N abundances have on the atmospheres of warm dwarfs, we fixed the [C/M] and [N/M] grids to solar values. The [C/Fe] and [N/Fe] values
reported for dwarfs in DR13 are calculated from windows after the best-fit parameters are determined, in similar fashion to the other individual elements. Figure~11 shows the improvement in the stellar parameters for members of the Pleiades star cluster. Some of the dwarfs in this young cluster are rotating with \vsini$>$ 50 km/s. In DR12, ASPCAP found a best-fit solution for metal-poor stars. The shallower the lines because of rotational broadening, the more metal-poor the star was reported to be. With the DR13 grid, there is no longer
a prominent trend in [M/H] with \vsini, and the mean value of the cluster is now at the expected metallicity.

\begin{figure*}[htb]
\begin{center}
\includegraphics[width=5in]{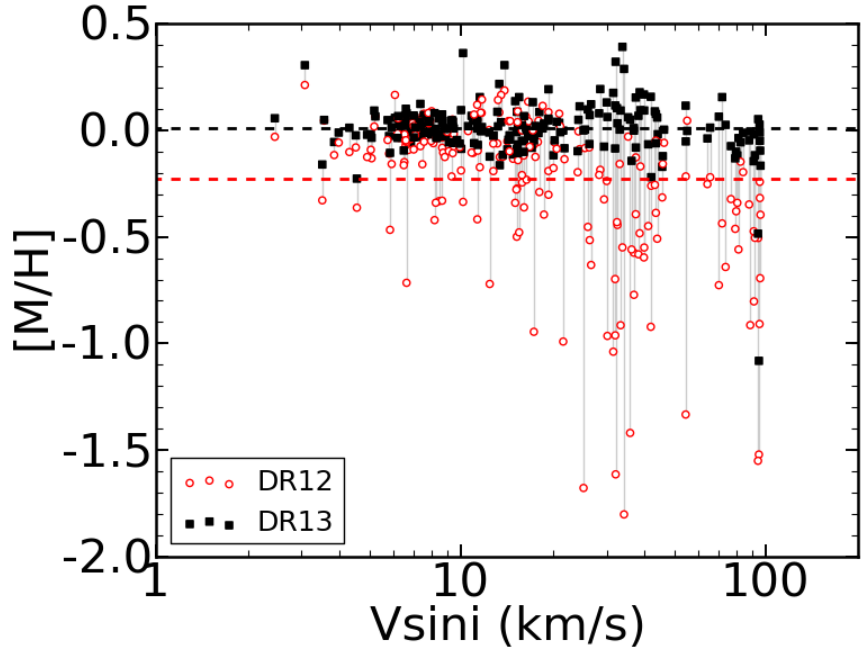}
\caption{A comparison of the [M/H] derived for Pleiades cluster dwarfs from DR12 (red open circles) and DR13 (black filled squares). The Pleiades has many rapidly spinning stars with rotationally broadened, shallow lines. When analyzed with the DR12 version of ASPCAP, which did not include rotational broadening, the best $\chi^2$ fit for the high \vsini{} stars was achieved for synthetic spectra with metallicity much lower than that known for the Pleiades. Therefore, the cluster average was sub-solar (red line). In DR13, where the dwarfs are run through a grid that includes a \vsini{} dimension, all Pleiades stars fall close to the correct value (black line), regardless of broadening.
}
\label{pleiades}
\end{center}
\end{figure*}

\subsection{Data Access}
Data access via the CAS and SAS is similar to that for DR12 \citep{holtz2015}; ``dr12'' should be replaced with ``dr13'' in the pathname or DR13 as the context in CASJobs. Some of the tag/column names have been modified. While raw abundances are still given in the FELEM arrays, ``calibrated” abundances are now presented in X\_H  and X\_M arrays
as well as in individual X\_FE tags/columns. For dwarfs, there is now a column in the allStar fits file on the SAS or in the aspcapStar table on the CAS that reports the \vsini. DR13 also includes a new catalog of red clump stars based on the new ASPCAP parameters, following the method of \citet{bovy2014}, available at the location described in Table 2.

\section{The Future}
SDSS-IV will continue to add to the SDSS legacy of data and data analysis tools in upcoming data releases. DR14 is scheduled for July 31, 2017, and there will be yearly data releases until the end of the survey in 2020. For MaNGA, 
future data releases will include more data cubes of galaxies that are currently being observed. DR14 will include
2744 additional galaxies observed by MaNGA. In DR15, a data interface system that includes web- and python-based tools to 
access a data analysis pipeline that carries out continuum and emission-line 
fitting to provide estimates of stellar and gas  kinematics, emission line fluxes, and absorption line index measurements
will be released. In addition, MaNGA has started a bright-time observing program, piggy-backing on APOGEE-2, to build a 
stellar library. These reduced optical stellar spectra will be included in DR15. For eBOSS, future 
data releases will provide sufficient redshifts of LRG, ELG, and quasars to be of cosmological interest, on its way to the 
goals described in \S1. For example, DR14 contains the spectra for 2480 square degrees observed by eBOSS. 
TDSS and SPIDERS will also release many more spectra, redshifts, and classifications for variable 
and X-ray emitting objects, respectively. For APOGEE-2, the next data release (DR14) will contain spectra for an 
additional 98,882 stars from the 
APOGEE-2N spectrograph at APO observed under SDSS-IV, and DR16 and subsequent DRs will provide spectra taken both with the 
APOGEE-2N and the APOGEE-2S spectrograph at LCO. In addition to observing red giants as the main target category, data 
will be obtained for RR Lyrae stars in the bulge, dwarf spheroidal galaxies, the Magellanic Clouds, {\it Kepler} Objects of 
Interest, and targets in the K2 fields.

\section{Acknowledgements}

We would like to thank the Center for Data Science (\url{http://cds.nyu.edu}) at New York
University for their hospitality during DocuFeest 2016.

Funding for the Sloan Digital Sky Survey IV has been provided by
the Alfred P. Sloan Foundation, the U.S. Department of Energy Office of
Science, and the Participating Institutions. SDSS-IV acknowledges
support and resources from the Center for High-Performance Computing at
the University of Utah. The SDSS web site is www.sdss.org.

SDSS-IV is managed by the Astrophysical Research Consortium for the 
Participating Institutions of the SDSS Collaboration including the 
Brazilian Participation Group, the Carnegie Institution for Science, 
Carnegie Mellon University, the Chilean Participation Group, the French Participation Group, Harvard-Smithsonian Center for Astrophysics, 
Instituto de Astrof\'isica de Canarias, The Johns Hopkins University, 
Kavli Institute for the Physics and Mathematics of the Universe (IPMU) / 
University of Tokyo, Lawrence Berkeley National Laboratory, 
Leibniz Institut f\"ur Astrophysik Potsdam (AIP),  
Max-Planck-Institut f\"ur Astronomie (MPIA Heidelberg), 
Max-Planck-Institut f\"ur Astrophysik (MPA Garching), 
Max-Planck-Institut f\"ur Extraterrestrische Physik (MPE), 
National Astronomical Observatory of China, New Mexico State University, 
New York University, University of Notre Dame, 
Observat\'ario Nacional / MCTI, The Ohio State University, 
Pennsylvania State University, Shanghai Astronomical Observatory, 
United Kingdom Participation Group,
Universidad Nacional Aut\'onoma de M\'exico, University of Arizona, 
University of Colorado Boulder, University of Oxford, University of Portsmouth, 
University of Utah, University of Virginia, University of Washington, University of Wisconsin, 
Vanderbilt University, and Yale University.

\bibliographystyle{apj}

\end{document}